\documentclass[preprintnumbers,amsmath,amssymb,superscriptaddress,prx,notitlepage,twocolumn,longbibliography]{revtex4-1}

\include{amsmath}
\usepackage{graphicx}
\usepackage{subfigure}
\usepackage{amsmath}
\usepackage{amsfonts}
\usepackage{color}
\usepackage[latin1]{inputenc}
\usepackage{times}

\newcommand{\therm}{\mathrm{th}}

\begin{document}

\title[]{Critical quantum fluctuations and photon antibunching in optomechanical systems\\ with large single-photon cooperativity}
\author{Kjetil B{\o}rkje}
\affiliation{Department of Science and Industry Systems, University of South-Eastern Norway, PO Box 235, Kongsberg, Norway}
\date{\today}

\begin{abstract}
A pertinent question in cavity optomechanics is whether reaching the regime of large single-photon cooperativity, where the single-photon coupling rate exceeds the geometric mean of the cavity and mechanical decay rates, can enable any new phenomena. We show that in some multimode optomechanical systems, the single-photon cooperativity can indeed be a figure of merit. We first study a system with one cavity mode and two mechanical oscillators which combines the concepts of levitated optomechanics and coherent scattering with standard dispersive optomechanics. Later, we study a more complicated setup comprising three cavity modes which does not rely on levitated optomechanics and only features dispersive optomechanical interactions with direct cavity driving. These systems can effectively realize the degenerate or the nondegenerate parametric oscillator models known from quantum optics, but in the unusual finite-size regime for the fundamental mode(s) when the single-photon cooperativity is large. We show that the response of these systems to a coherent optical probe can be highly nonlinear in probe power even for average photon occupation numbers below unity. The nonlinear optomechanical interaction has the peculiar consequence that the probe drive will effectively amplitude-squeeze itself. For large single-photon cooperativity, this occurs for small occupation numbers, which enables observation of nonclassical antibunching of the transmitted probe photons due to a destructive interference effect. Finally, we show that as the probe power is increased even further, the system enters a critical regime characterized by intrinsically nonlinear dynamics and non-Gaussian states.
\end{abstract}

\maketitle

\section{Introduction}
In single-mode dispersive cavity optomechanics \cite{Aspelmeyer2014RMP}, the oscillatory motion of a mechanical element modulates the resonance frequency of an electromagnetic cavity mode. This simple interaction has enabled quantum ground state cooling \cite{Teufel2011Nature,Chan2011Nature} and squeezing \cite{Pirkkalainen2015PRL,Lecocq2015PRX,Wollman2015Science} of a motional mode of micromechanical objects consisting of macroscopic numbers of atoms. It has also provided a new way to manipulate the quantum noise properties of light \cite{Brooks2012Nature,Purdy2013PRX,Safavi-Naeini2013Nature} and to observe quantum correlations between radiation and motion \cite{Purdy2013Science,Palomaki2013Science}. Extending the system to several mechanical and optical modes has even made it possible to entangle mechanical modes of remote physical objects \cite{Riedinger2018Nature,Ockeloen-Korppi2018Nature}. 

The radiation pressure interaction between motion and radiation is characterized by a single-photon coupling rate $g_0$, which is the cavity frequency shift caused by the size of the quantum zero-point motion of the mechanical oscillator. In all experiments to date, this coupling rate is orders of magnitude smaller than the linewidth $\kappa$ of the cavity mode. However, coherent driving of the cavity mode will produce an effective linear coupling between mechanical and optical fluctuations with a coupling rate $g_0 \sqrt{n_\mathrm{pht}}$, where $n_\mathrm{pht}$ is the average number of photons in the cavity. This enables reaching the regimes of large cooperativity \cite{Teufel2011Nature,Chan2011Nature,Weis2010Science}, strong coupling \cite{Groblacher2009Nature,Teufel2011Nature_2}, and even ultrastrong coupling \cite{Peterson2019PRL}. 

The enhanced coupling rate nevertheless comes at the price of linearized dynamics such that the optomechanical system in some sense behaves more classically. One way to appreciate this is to note that $g_0  \propto \sqrt{\hbar}$ whereas $n_\mathrm{pht} \propto \hbar^{-1}$, such that the enhanced coupling rate is independent of $\hbar$ \cite{Aspelmeyer2014RMP}. Observing quantum effects then requires some kind of quantitative comparison, e.g., to rule out classical noise \cite{Jayich2012NJP,Safavi-Naeini2013NJP}. Alternatively, one needs to drive the system with quantum states or take advantage of measurement-induced nonlinearities \cite{Hong2017Science,Riedinger2018Nature}.

Effects due to the intrinsic nonlinear interaction between mechanical and optical fluctuations are generally expected to become relevant in the single-photon strong coupling regime $g_0 \gtrsim \kappa$ \cite{Rabl2011PRL,Nunnenkamp2011PRL,Ludwig2012PRL,Stannigel2012PRL,Borkje2013PRL,Lemonde2013PRL,Kronwald2013PRL,Liao2016PRL}. We note, however, that some exceptions to this requirement have been predicted in cases of optical \cite{Lu2015PRL} or mechanical \cite{Lemonde2016NatComm} parametric driving, in systems driven close to an instability \cite{Lu2013SciRep,Xu2015PRA}, or in carefully designed multimode systems \cite{Xuereb2012PRL,Li2016JOpt}.

Although reaching the single-photon strong coupling regime of optomechanics is difficult, there is great experimental progress on reducing the mechanical dissipation rates in optomechanical systems. Quality factors $Q_m \gtrsim 10^8$ have been reported for flexural modes in dielectric membranes \cite{Reinhardt2016PRX,Norte2016PRL,Tsaturyan2017NatNano,Reetz2019PRApplied} or nanobeams \cite{Ghadimi2018Science}, and localized acoustic modes in suspended photonic crystals can have quality factors as large as $Q_m \sim 10^{10}$ \cite{Maccabe2019}. In light of this, one may wonder whether any new phenomena can be realized in the regime where the single-photon optomechanical cooperativity
\begin{equation}
\label{eq:SinglePhotCoopDef}
{\cal C}_0 = \frac{4 g_0^2}{\kappa \gamma_1} 
\end{equation}
exceeds unity, where $\gamma_1$ is the mechanical decay rate. Several experiments have in fact reached this regime. Most of them are in the unresolved sideband regime where the cavity decay rate $\kappa$ is much larger than the mechanical resonance frequency \cite{Wilson2015Nature,Leijssen2017NatComm,Guo2019,Fogliano2019,Zoepfl2019}. However, a value of $C_0 \sim 8$ has been reported for a trampoline membrane-in-the-middle setup where the mechanical resonance frequency and the cavity decay rate were comparable \cite{Reinhardt2016PRX}. 

To our knowledge, the single-photon cooperativity \eqref{eq:SinglePhotCoopDef} does not play any significant role in the single-mode optomechanical system as long as $g_0 \ll \kappa$. The relevant figure of merit is in fact the enhanced cooperativity ${\cal C} = {\cal C}_0 n_\mathrm{pht}$. A large ${\cal C}_0 $ can of course be advantageous since it requires less circulating power in the cavity mode, but it does not bring on any new phenomena. An exception would be if the mechanical and cavity resonance frequencies are comparable such that two-photon creation/annihilation processes become relevant \cite{Jansen2019PRB}, but this is typically far from the case. 

In this paper, we aim to study the simplest multimode optomechanical system where the single-photon cooperativity ${\cal C}_0$ is a genuine figure of merit. To do this, we will have to assume experimental setups beyond what has already been realized, but we will restrict ourselves to the weak coupling limit $g_0 \ll \kappa$. Unlike previous proposals for realizing nonlinear effects in the weak coupling limit \cite{Lu2015PRL,Lemonde2016NatComm,Lu2013SciRep,Xu2015PRA,Xuereb2012PRL,Li2016JOpt}, our proposal benefits from the smallness of the mechanical decay rate $\gamma_1$. The purpose of our study is to stimulate further work, both theoretical and experimental, towards bringing cavity optomechanics into the nonlinear regime. 

We will start by studying a system where one cavity mode couples to two mechanical oscillators, one of which is the motion of a nanoparticle levitated by an optical tweezer. Optically levitated nanoparticles have recently been cooled to the motional ground state \cite{Delic2019} by the so-called coherent scattering technique \cite{Vuletic2001PRA,Delic2019PRL,Windey2019PRL,Gonzales-Ballestero2019PRA}. We consider such a setup where tweezer photons can scatter into an undriven cavity mode due to the motion of a nanoparticle. The system is thus distinct from that of dispersive optomechanics of a driven cavity with two mechanical oscillators \cite{Massel2012NatComm,Shkarin2014PRL} and we will explain why this is important. Nevertheless, since dispersive and driven optomechanics is more common in experiments, we also show that our model can be realized in such systems as well. However, we argue that this might require complicated setups involving three cavity modes.

The multimode systems we study will be shown to effectively realize the degenerate or the nondegenerate parametric oscillator models known from quantum optics with nonlinear media \cite{Walls2008Book,CarmichaelBook2008,McNeil1978OptComm,Drummond1980OpticaActa}. We will see that for sufficiently large single-photon cooperativity ${\cal C}_0$, these models are realized in the so-called finite-size regime for the fundamental mode \cite{CarmichaelBook2008,Veits1997PRA,Veits1997PRA_2,Veits1995PRA}. This is an unusual regime in the context of nonlinear media and has not been studied in much detail. 

After showing how our model maps onto the effective parametric oscillator models, we study the reponse of the optomechanical system to an optical coherent probe drive. This is equivalent to studying second harmonic generation in the effective models. We will see that in the regime of large ${\cal C}_0$, the critical behaviour known from mean-field theory \cite{McNeil1978OptComm,Drummond1980OpticaActa} is smeared out, predominantly by quantum fluctuations. For sufficiently large probe drive, we show that the system reaches a critical regime characterized by nonlinear interactions between optical and mechanical fluctuations.

We also show that the system has a highly nonlinear response to the optical probe drive in a narrow frequency window, which is a nonlinear version of optomechanically induced transparency \cite{Agarwal2010PRA,Weis2010Science}. Similar effects have been studied in single-mode optomechanics \cite{Borkje2013PRL,Lemonde2013PRL,Kronwald2013PRL}, but unlike in that case, the effect is here not limited by the smallness of $g_0/\kappa$. Due to the nonlinear interaction, the probe drive will also tend to amplitude-squeeze itself. For large ${\cal C}_0$, this autonomous squeezing is significant even in the regime of cavity occupation numbers well below unity, and we will see that it facilitates the observation of photon antibunching. 

This paper is organized as follows. In Section \ref{sec:Model1}, we describe the optomechanical model which involves the levitated nanoparticle. We then define the normal modes of the system in Section \ref{sec:NormModes} and derive a master equation expressed in terms of these normal modes in Section \ref{sec:EffModel}. In Section \ref{sec:NonlinInt}, we express the nonlinear part of the optomechanical interaction in terms of normal modes, where we recognize the effective parametric oscillator models. We first study the steady-state of the undriven effective models in Section \ref{sec:Cooling}, before we consider the response to an optical probe drive in Section \ref{sec:SHD}. In Section \ref{sec:Impl}, we show how our model can also be realized in multimode, driven, and dispersive optomechanics. Finally, we summarize and discuss future possible directions in Section \ref{sec:Disc}.

\section{Model}
\label{sec:Model1}

\begin{center}
\begin{figure}[htb]
\includegraphics[width=.45\textwidth]{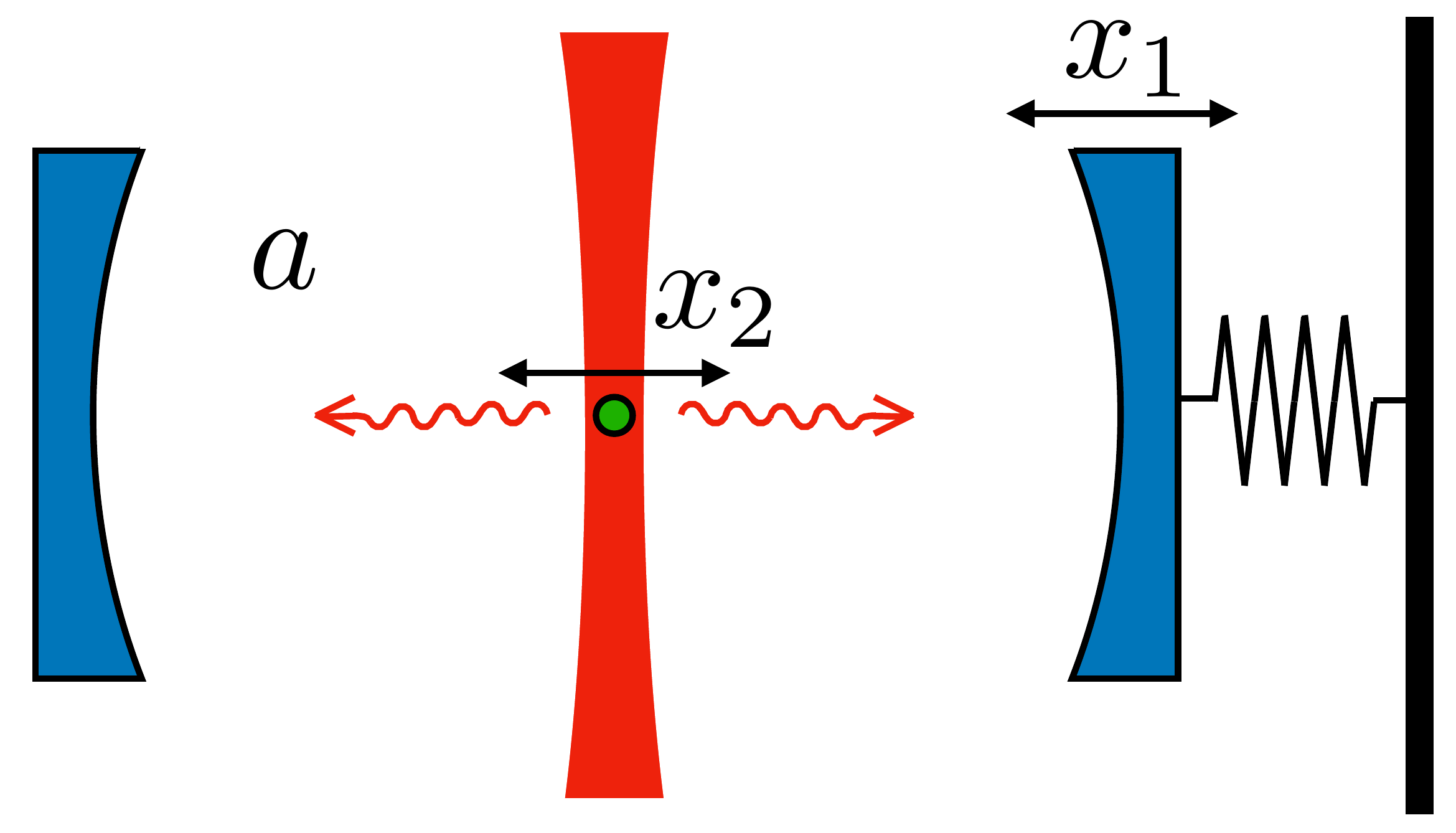}
\caption{Schematic of the setup. A nanoparticle (green) is levitated by an optical tweezer (red) and placed inside an optical cavity with a movable mirror. Alternatively, one could consider both mirrors fixed and include a dielectric membrane inside the cavity. The undulating red arrows indicate that photons can scatter from the tweezer beam and into the cavity mode due to interaction with the nanoparticle.}
\label{fig:CSModel}
\end{figure}
\end{center}

We consider the system shown in Figure \ref{fig:CSModel}, where a nanoparticle levitated by an optical tweezer is placed inside an optical cavity. The motion of the particle in the electromagnetic trap can then cause scattering of tweezer photons into a cavity mode. In addition to the levitated nanoparticle, we also assume that the cavity mode is influenced by another mechanical oscillator. This is depicted as a movable end mirror in Figure \ref{fig:CSModel}, but one could also imagine implementations with other types of oscillators, e.g., membrane-in-the-middle \cite{Thompson2008Nature,Reinhardt2016PRX,Norte2016PRL,Tsaturyan2017NatNano}. 

Note that we do not consider separate laser driving of the optical cavity here, which means that all the photons that enter the cavity originate from the trapping field and have scattered off of the nanoparticle. 

\subsection{System Hamiltonian}
\label{sec:SysHam}

We will consider a setup where the trap laser is polarized perpendicular to the cavity axis and where the nanoparticle is positioned at a node of the optical cavity mode. In this case, any scattering of photons from the trap laser into the optical cavity (and vice versa) must be caused by the {\it motion} of the nanoparticle \cite{Delic2019PRL,Gonzales-Ballestero2019PRA}. In other words, there would be no direct scattering from a static particle. Furthermore, the scattering of photons into the cavity mode is only caused by motion along the cavity axis \cite{Delic2019PRL,Gonzales-Ballestero2019PRA}. For this reason, we need only consider the motion of the nanoparticle along this axis, i.e., we can treat it as a one-dimensional oscillator. We will also only consider a single mode of the movable mirror's motion, as well as a single optical cavity mode.

The Hamiltonian is defined as $H = H_\mathrm{free} + H_\mathrm{int,1} + H_\mathrm{int,2} $ where
\begin{equation}
\label{eq:Hfree}
H_\mathrm{free} =   - \hbar \Delta a^\dagger a +  \hbar \omega_{m,1} b^\dagger_1 b_1 + \hbar \omega_{m,2} b^\dagger_2 b_2 
\end{equation}
contains the free harmonic oscillator Hamiltonians for the cavity mode with photon annihilation operator $a$, a mechanical mode of the mirror's motion with phonon annihilation operator $b_1$, and a mechanical mode of the nanoparticle's motion with phonon annihilation operator $b_2$. The detuning $\Delta = \omega_\mathrm{tw} - \omega_c$ is the difference between the optical trapping laser frequency $\omega_\mathrm{tw}$ and the cavity resonance frequency $\omega_c$, and the mechanical oscillators have resonance frequencies $\omega_{m,j}$, $j = 1,2$. We assume $  \omega_{m,1} , \omega_{m,2} \ll \omega_\mathrm{tw}, \omega_c$.

We will also assume that the mirror mode frequency is comparable to twice the nanoparticle mode frequency. Specifically, we require
\begin{equation}
\label{eq:RoughFreqRel}
\left|\omega_{m,1} - 2\omega_{m,2} \right| \ll \omega_{m,2} .
\end{equation}
We emphasize that there is no need for fine-tuning the relation between the two mechanical frequencies, as long as \eqref{eq:RoughFreqRel} is satisfied. We note that the nanoparticle's resonance frequency can to some extent be tunable \cite{Delic2019PRL,Gonzales-Ballestero2019PRA}. In addition, a mirror or a membrane-in-the-middle will have several mechanical modes that can couple to the same cavity mode. These properties should make it feasible to meet the requirement \eqref{eq:RoughFreqRel}.

The first part of the interaction Hamiltonian is
\begin{equation}
\label{eq:Hint1}
H_\mathrm{int,1}  = \hbar g_{0} x_1 a^\dagger a   .
\end{equation}
We let $x_j = b_j + b_j^\dagger$ denote the position operator for oscillator $j$ in units of its zero point motion. This is the standard radiation pressure interaction between the movable mirror and the optical cavity. As already mentioned, we will assume that the single-photon coupling rate $g_0$ is much smaller than the decay rate of the optical cavity.

The second part of the interaction Hamiltonian is
\begin{equation}
\label{eq:Hint2}
H_\mathrm{int,2}  =  \hbar G x_2 \left( a + a^\dagger \right) .
\end{equation}
This describes the interaction between the optical tweezer field, the optical cavity field, and the motion of the levitated nanoparticle, for the particular positioning and polarization described above. It originates from the interference term between the tweezer and the cavity field in the Hamiltonian of the electromagnetic field, where the tweezer field has been approximated by its average value. The coupling rate $G$ depends on a number of parameters \cite{Delic2019PRL,Gonzales-Ballestero2019PRA}. Most importantly, it is proportional to the square root of the laser power of the tweezer and thus tunable. The term $\sim b^{(\dagger)}_2 a^\dagger$ describes a process where a tweezer photon scatters into the optical cavity mode $a$ while simultaneously annihilating (creating) a phonon in the nanoparticle mode $b_2$. We will assume below that it is possible to reach the (many-photon) strong coupling regime where $G$ exceeds the cavity decay rate. We neglect interaction terms of higher order between modes $a$ and $b_2$, since we are working in the limit of weak single-photon coupling.
 
We emphasize that the system we study is {\it distinct} from simply having two mechanical oscillators couple to a driven cavity mode. This is clear from the absence of a term $H' = \hbar G x_1 (a + a^\dagger)$ in the interaction Hamiltonian \eqref{eq:Hint1}. In a standard and coherently-driven optomechanical system, such a term originates from \eqref{eq:Hint1} when displacing the operator $a \rightarrow a + \alpha$ by its coherent amplitude $\alpha$. The lack of this usual 'linearized' optomechanical interaction for mechanical mode $b_1$ in this case is due to the fact that we are not coherently driving the cavity and that there is no direct scattering into the cavity mode with our assumptions, only scattering caused by the nanoparticle's motion. 

The absence of $H'$ will be crucial to realize the effects we study here. The reason is that for a red-detuned drive ($\Delta < 0$), such a term would lead to the well-known optomechanical damping of mode $b_1$. Even if the detuning $\Delta$ is far away from the optimal damping condition, the mechanical linewidth broadening can be significant for sufficiently large laser power. We wish, however, to preserve the narrow linewidth of mode $b_1$. This is why we have designed the system such that modes $a$ and $b_1$ only interact through the intrinsic, nonlinear radiation pressure interaction. We will comment further on this issue in Section \ref{sec:EffCoop}.

\subsection{Dissipation}
\label{sec:Diss}

We now describe the interaction of the cavity and mechanical modes with their respective environments. While this interaction is of the standard form, the strong coupling between the oscillator $b_2$ and the cavity mode $a$ will give rise to unusual terms in the effective master equation describing the system \cite{Lemonde2015PRA}. To appreciate the physical origin of these effects, we therefore include some details on how the interaction with the environment is incorporated in the effective description of the system. 

For simplicity, we consider the cavity to be one-sided, i.e., its only decay channel is to the electromagnetic continuum through one mirror (the left mirror in Figure \ref{fig:CSModel}). This assumption is not crucial to our results and the model can be straightforwardly generalized to include other decay channels. The external electromagnetic modes, and their coupling to the cavity mode, are included by extending the Hamiltonian with $H_{\mathrm{ext},c} =   \hbar \sum_k  \omega_{c,k} f^\dagger_k f_k + H_{\mathrm{s-b},c}$, where the system-bath interaction is
\begin{equation}
\label{eq:HsbCav}
H_{\mathrm{s-b},c} =    \sum_k  \hbar \lambda_{c,k} \left( f_k^\dagger a + a^\dagger f_k \right)  .
\end{equation}
We have defined $f_k$ as the photon annihilation operator for an external mode with frequency $\omega_{c,k}$. This is expressed in terms of discrete modes labeled by the integer $k$, but we will later take the continuum limit. Equation \eqref{eq:HsbCav} describes the bilinear interaction between the cavity mode and the outside modes, where $\lambda_{c,k}$ are coupling constants. We have neglected two-photon creation (annihilation) processes $f_k^\dagger a^\dagger$ ($f_k a$), as these are off-resonant by a frequency $\sim 2 \omega_c$ and thus strongly suppressed. To include dissipation properly in our setup, it is crucial to note that when $a, f_k$ refer to the frame rotating at the laser frequency, the mode frequencies $\omega_{c,k}$ can be negative \cite{Lemonde2015PRA}.  

The cavity mode will in practice only interact with outside modes in a narrow frequency interval around $\omega_c$ of width $\ll \omega_c$. We may then approximate the coupling rates $\lambda_{c,k} \rightarrow \lambda_c$ and the bath density of states $\rho_c$ by constants in the frequency interval of interest. This is equivalent to treating the electromagnetic environment as a Markovian bath \cite{Clerk2010RMP}. In the absence of optomechanical interaction, a single photon Fock state in the cavity will then decay at a rate 
\begin{equation}
\label{eq:kappaDef}
\kappa = 2\pi \rho_c \lambda_c^2
\end{equation}
due to emission into the outside modes. We will refer to $\kappa$ as the cavity decay rate in the following. We assume $\omega_{m,2} \gg \kappa$, such that both mechanical modes are in the resolved sideband regime. 

The interaction between the mechanical modes and their environment is described in a similar way. This involves adding $H_{\mathrm{ext},j} =    \sum_k \hbar \omega_{j,k} g^\dagger_{j,k} g_{j,k} + H_{\mathrm{s-b},j}$ with  
\begin{equation}
\label{eq:HsbMech}
H_{\mathrm{s-b},j} =    \sum_k  \hbar \lambda_{j,k} \left( g_{j,k} + g_{j,k}^\dagger \right) x_j  
\end{equation}
to the total Hamiltonian, for both $j = 1,2$. The environmental mode frequencies $\omega_{j,k}$ are now strictly positive. We do not perform the rotating wave approximation at this point, since the mechanical mode $b_2$ may have support at negative frequencies for a sufficiently large coupling rate $G$ \cite{Lemonde2015PRA}. 

For simplicity, we will again apply the Markov approximation by replacing the coupling rates $\lambda_{j,k} \rightarrow \lambda_{j} $ by constants, and assuming constant densities of states $\rho_{j}$ in the frequency intervals of interest. This approximation may be less accurate than for the cavity environment if $G$ becomes comparable to $\omega_{m,2}$, since the relevant frequency interval relative to the absolute frequency scale $\omega_{m,2}$ is then larger \cite{Lemonde2015PRA}. However, the approximation may still be fairly good in the limit $G \ll \omega_{m,2}$ that we consider below. More importantly, we do not expect corrections to these approximations to affect our conclusions in any significant way. 

In the absence of optomechanical interactions, a single phonon Fock state in the mechanical mode $b_j$ will decay at a rate 
\begin{equation}
\label{eq:gammaDef}
\gamma_j = 2\pi \rho_{j} \lambda_{j}^2 .
\end{equation}
We will refer to $\gamma_j$ as the intrinsic mechanical decay rate of mode $b_j$. 

We assume $\hbar \omega_c \gg k_B T$, where $T$ is temperature. The unperturbed optical bath modes are then in the vacuum state $\langle f^\dagger_k f_{k'}\rangle = 0$. The mechanical bath modes may on the other hand be thermally occupied, such that $\langle g^\dagger_{j,k} g_{j,k'} \rangle = n_B(\omega_{j,k}) \delta_{k,k'}$ where
\begin{equation}
\label{eq:nPDef}
n_B(\omega) = \frac{1}{e^{\hbar \omega/(k_B T)} - 1}
\end{equation}
is the Planck distribution. We will in the following assume
\begin{equation}
\label{eq:ThermalCond}
\gamma_j n_B(\omega) \ll \kappa , 
\end{equation}
for $\omega$ on the order of $\omega_{m,1}, \omega_{m,2}$. Physically, this means that the rate at which excitations enter the system due to thermal bath phonons is much smaller than the rate at which they decay through the cavity mirror. This is an experimentally relevant assumption and is a prerequisite for optomechanical sideband cooling to the motional ground state \cite{Teufel2011Nature,Chan2011Nature,Delic2019}.

\section{Normal modes}
\label{sec:NormModes}

In order to derive an effective model for our setup that properly includes dissipation, we must first define the normal modes of the system and express the system-bath interaction in terms of these. 

\subsection{Diagonalization of bilinear terms}

We choose the laser to be red-detuned with respect to the cavity. Specifically, we let 
\begin{align}
\label{eq:deltaDef}
\Delta & = -(\omega_{m,2} + \delta), 
\end{align}
with $|\delta| \ll \omega_{m,2}$. This means that the cavity mode $a$ and the mechanical mode $b_2$ are degenerate (for $\delta = 0$) or almost degenerate (for $\delta \neq 0$). We will consider an effective coupling rate $G$ that exceeds the decay rates $\kappa , \gamma_2$ of the individual modes. We note that this regime of {\it linear} strong coupling has been reached in various experimental implementations of dispersive optomechanics, leading to normal-mode splitting \cite{Marquardt2007PRL,Groblacher2009Nature,Teufel2011Nature_2}. The normal modes, which are the long-lived excitations of the system, can be thought of as hybrids of photons and phonons. 

In general, the operators $a,b_2$ and the annihilation operators $c_\pm$ of the normal modes are related by a symplectic transformation. This transformation can be somewhat unwieldy, in particular for $\delta \neq 0$. However, since we will consider coupling rates and detunings $G, |\delta| \ll \omega_{m,2}$, we can calculate the transformation perturbatively in $G/\omega_{m,2} , |\delta|/\omega_{m,2}$. For convenience, we define the parameters
\begin{align}
\label{eq:qpDef}
q_\pm & = \frac{rG}{(1+r^2) \omega_{\pm,0}}  , \\
 p & = \frac{r\delta}{(1+r^2) (\omega_{-,0} + \omega_{+,0})} ,
\end{align}
with
\begin{equation}
\label{eq:rDef}
r = \frac{2G/\delta}{1 + \text{sgn}(\delta) \sqrt{1 + \left(2G/\delta\right)^2}} 
\end{equation}
and 
\begin{equation}
\label{eq:pol0freq}
\omega_{\pm,0} = \omega_{m,2} + \frac{\delta}{2}  \pm \sqrt{ \left(\frac{\delta}{2}\right)^2 + G^2} .
\end{equation}
Note that $\omega_{\pm,0}$ are the normal mode resonance frequencies one would find in the rotating wave approximation $x_2 (a+ a^\dagger) \approx b_2^\dagger a+ a^\dagger b_2$. We will, however, go beyond that approximation here.

To second order in $q_\pm,p$, the transformation to bosonic normal mode operators $c_\pm$ is given by
\begin{align}
\label{eq:symplectic}
a & = \frac{1 + \frac{1}{2} \left(q_+^2 + p^2\right) }{\sqrt{1+r^2}} c_+ - \frac{q_+ - rp}{\sqrt{1+r^2}} c_+^\dagger \notag \\
& - \frac{r \left[ 1 + \frac{1}{2} \left(q_-^2 + p^2\right) \right]}{\sqrt{1+r^2}} c_-  - \frac{p + rq_- }{\sqrt{1+r^2}} c_-^\dagger   \\
b_2 & = \frac{1 + \frac{1}{2} \left(q_-^2 + p^2\right)}{\sqrt{1+r^2}} c_- + \frac{q_- - rp}{\sqrt{1+r^2}} c_-^\dagger \notag 
\\ 
& + \frac{r \left[ 1 + \frac{1}{2} \left(q_+^2 + p^2\right) \right]}{\sqrt{1+r^2}} c_+ - \frac{p + r q_+}{\sqrt{1+r^2}} c_+^\dagger .
\end{align}
We will limit ourselves to $|\delta| \lesssim G$, in which case the parameter $r$ is of order 1. In this regime, the mechanical and optical modes are always strongly hybridized.

In terms of the normal mode operators, we can now write the bilinear part of the system Hamiltonian as 
\begin{equation}
\label{eq:HQuadDiag}
H_\mathrm{free} + H_{\mathrm{int},2} = \hbar \omega_+ c_+^\dagger c_+ +  \hbar \omega_- c_-^\dagger c_-  + \hbar \omega_{m,1} b_1^\dagger b_1. 
\end{equation}
The normal mode resonance frequencies are
\begin{equation}
\label{eq:polfreq}
\omega_{\pm} = \omega_{\pm,0} - \frac{r^2 (4 G^2 + \delta^2)}{2(1+r^2)^2 \omega_{m,2}} 
\end{equation}
to second order in $G/\omega_{m,2}, |\delta|/\omega_{m,2}$. From Equation \eqref{eq:pol0freq}, we note that the the normal mode frequency splitting depends on $G$ and can thus be controlled by adjusting the tweezer laser power. 

\subsection{Dissipation}

We may now express the system-bath interaction Hamiltonians in terms of the normal mode operators $c_\pm$. To first order in $q_\pm,p$, the interaction with the optical bath in Equation \eqref{eq:HsbCav} becomes
\begin{align}
\label{eq:HsbCav2}
H_{\mathrm{s-b},c} & =   
 \hbar \sqrt{\frac{\kappa}{2\pi \rho_c (1+r^2)}} \sum_k \Big\{ f_k^\dagger c_+  - r f_k^\dagger c_- \\
& - (q_+ - rp) f_k^\dagger c_+^\dagger  - (p + rq_-) f_k^\dagger c_-^\dagger  + \text{h.c.}    \Big\} \notag . 
\end{align}
The first line is similar to the original system-bath interaction in Equation \eqref{eq:HsbCav} and describes processes where normal mode particles are destroyed and photons are created in the optical bath. 

\begin{center}
\begin{figure}[htb]
\includegraphics[width=.45\textwidth]{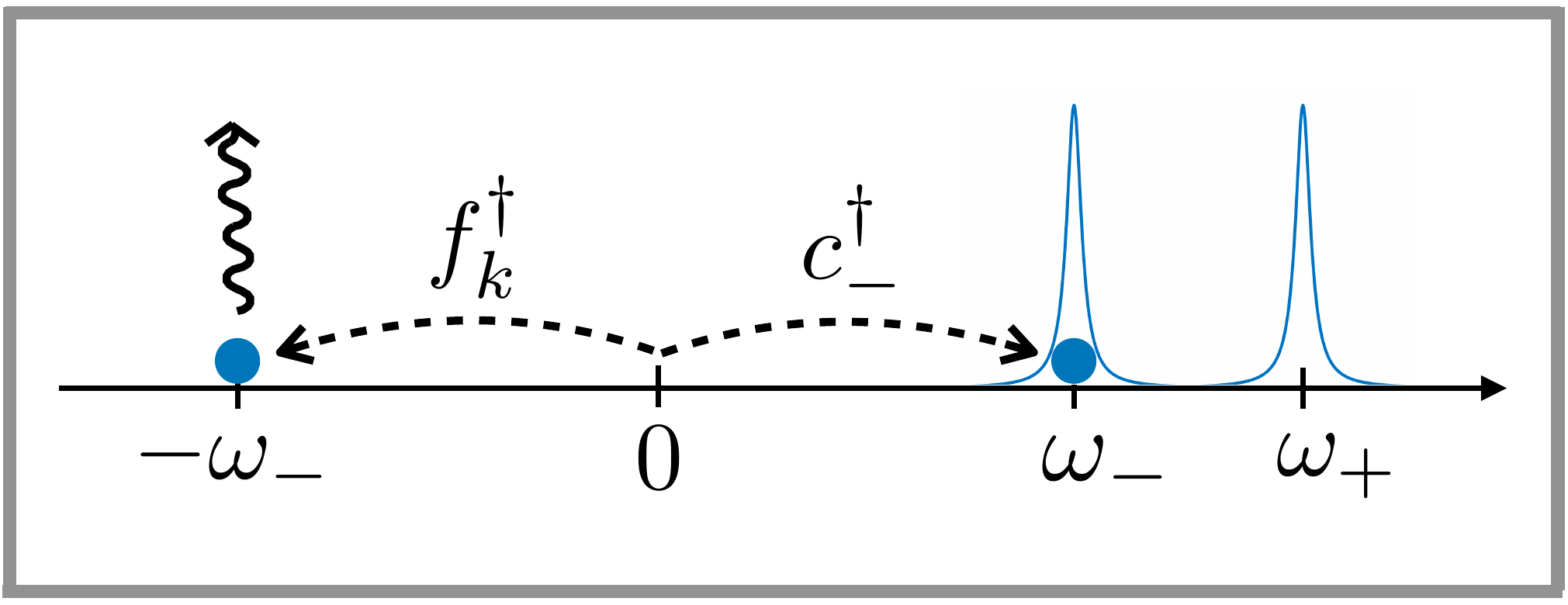}
\caption{Illustration of quantum heating. In a dissipative pair creation process, a photon is emitted into the bath and a particle is created in normal mode $c_-$. Such processes contribute to an effective nonzero thermal occupation of the normal modes.}
\label{fig:QuantHeat}
\end{figure}
\end{center}

The second line in \eqref{eq:HsbCav2}, on the other hand, describes processes where both normal mode particles and bath photons are created. These processes can also be resonant, as there are modes $f_k$ with negative resonance frequencies. Such a process is illustrated in Figure \ref{fig:QuantHeat}.
The consequence is that even if the bath modes $f_k$ are originally in the vacuum state, dissipation can lead to nonzero occupation of the normal modes. This effect has been referred to as quantum heating \cite{Lemonde2015PRA} and is related to the concept of quantum activation \cite{Marthaler2006PRA}. We note that even though it is useful to think of this as pairs of particles created from vacuum, the actual origin of all photons in this model is the trap laser. This is because the optomechanical interactions conserve photon numbers as long as $\omega_{m,1} , \omega_{m,2} \ll \omega_c$. These two-particle creation processes can be thought of as strong-coupling generalizations of correlated red- and blue-shifted photons emitted from weakly coupled optomechanical systems \cite{Borkje2011PRL}.

The interaction with the mechanical bath of mode $b_2$ in \eqref{eq:HsbMech}, with $j = 2$, becomes
\begin{align}
\label{eq:HsbMech2}
H_{\mathrm{s-b},2} &  =   
 \hbar \sqrt{\frac{\gamma_2}{2\pi \rho_2 (1+r^2)}} \sum_k \Big\{  \left[r\left(1 - q_+\right) - p\right] g_{2,k}^\dagger c_+ \notag \\
 & + \left(1 + q_- - rp\right) g_{2,k}^\dagger c_-   + \text{h.c.}    \Big\}  , 
\end{align}
again to first order in $q_\pm, p$. Here, we have performed the rotating wave approximation after expressing the system-bath interaction in terms of normal mode operators \cite{Lemonde2015PRA}. The rationale for this is that the modes $g_{2,k}$ all have positive resonance frequencies, such that two-particle creation terms are off-resonant by $\gtrsim \omega_{m,2}$.

\section{Effective model}
\label{sec:EffModel}

\subsection{Quantum Langevin equations}

We can now derive quantum Langevin equations for the normal mode operators $c_\pm$ by using input-output theory \cite{Gardiner1985PRA,Clerk2010RMP}. To do this, we define the optical bath input noise
\begin{equation}
\label{eq:InputOpt}
\xi(t) = -\frac{i}{\sqrt{2\pi \rho_c}} \sum_k e^{-i\omega_{c,k}(t-t_0)} f_k(t_0) , 
\end{equation}
and the mechanical bath input noise
\begin{equation}
\label{eq:InputOpt2}
\eta_2(t) = -\frac{i}{\sqrt{2\pi \rho_2}} \sum_k e^{-i\omega_{2,k}(t-t_0)} g_{2,k}(t_0) ,
\end{equation}
where $t_0$ is a time in the distant past. The occupation numbers of the bath modes at time $t_0$ are assumed to be the unperturbed ones, i.e., $\langle f^\dagger_k(t_0) f_{k'}(t_0) \rangle = 0 $ and $\langle g^\dagger_{2,k}(t_0) g_{2,k'}(t_0) \rangle = n_B(\omega_{2,k}) \delta_{k,k'}$. 

The quantum Langevin equations then become
\begin{align}
\label{eq:QLP}
\dot{c}_+ & = - \left(\frac{\kappa_+}{2} + i \omega_+\right) c_+ + \frac{\kappa_{+-}}{2} c_- + \frac{\tilde{\kappa}_{+-}}{2} c^\dagger_- \notag \\
& + \sqrt{\kappa} \, \frac{\xi + (q_+ - rp)  \xi^\dagger}{\sqrt{1+r^2}}  + \sqrt{\gamma_2} \, \frac{r(1-q_+) - p}{\sqrt{1+r^2}} \eta_2 \notag \\
& + \frac{i}{\hbar} \left[H_{\mathrm{int},1} , c_+\right] 
\end{align}
and
\begin{align}
\label{eq:QLM}
\dot{c}_- & = - \left(\frac{\kappa_-}{2} + i \omega_-\right) c_- + \frac{\kappa_{+-}}{2} c_+ - \frac{\tilde{\kappa}_{+-}}{2} c^\dagger_+ \notag \\
& - \sqrt{\kappa} \, \frac{r \xi - (p + r q_- )  \xi^\dagger}{\sqrt{1+r^2}}  + \sqrt{\gamma_2} \, \frac{1+ q_- - rp}{\sqrt{1+r^2}} \eta_2 \notag \\
& + \frac{i}{\hbar} \left[H_{\mathrm{int},1} , c_-\right] ,
\end{align}
where we have exploited $q_\pm , |p| \ll 1$. We have introduced the normal mode decay rates
\begin{align}
\label{eq:NormModDR}
\kappa_+ & = \frac{\kappa + r^2 \gamma_2}{1 + r^2} \approx \frac{\kappa}{1+r^2} \\
\kappa_- & =  \frac{r^2 \kappa + \gamma_2}{1 + r^2} \approx \frac{r^2 \kappa}{1+r^2} ,
\end{align}
which are both on the order of the cavity decay rate $\kappa$ with our assumptions $\gamma_2 \ll \kappa$, $r \sim {\cal O}(1)$. This means that the normal mode particles predominantly decay as photons. For $\delta = 0$, i.e., $r = 1$, the decay rates are $\kappa_\pm \approx \kappa/2$. 

The quantum Langevin equations also contain dissipative terms that are off-diagonal in normal mode index, governed by the parameter $\kappa_{+-} = r(\kappa - \gamma_2)/(1 + r^2)$. There are also two-mode dissipative squeezing terms proportional to $\tilde{\kappa}_{+-} = [ p + r (q_+ + q_-)/(1+r^2)] \kappa $. However, in the regime $\kappa \ll |\omega_+ - \omega_-| \ll \omega_{m,2}$ we consider here, these unconventional dissipative terms will be off-resonant, and we thus neglect them in the following.

\subsection{Effective master equation}

Let us now for a moment ignore the presence of mechanical mode $b_1$, i.e., remove the last term in Equations \eqref{eq:QLP} and \eqref{eq:QLM}. Calculating the average occupation numbers of the normal modes $n_{\therm,\pm} \equiv \langle c_\pm^\dagger c_\pm \rangle_0$, where the subscript $0$ indicates absence of interaction with mode $b_1$, then gives
 \begin{align}
\label{eq:ThermNorm}
n_{\therm,\pm} &  \approx \frac{r^{\pm 2} \gamma_2 n_B(\omega_\pm)}{\kappa} + \frac{r^2 \left(4G^2 + r^{\pm 2} \delta^2\right)}{4(1+r^2)^2 \omega_{m,2}^2} .
\end{align}
The first term comes from coupling to the thermal bath of oscillator $b_2$. The second term originates from the pair creation terms in Equation \eqref{eq:HsbCav2}, i.e., the quantum heating effect. 

Note that with our assumptions $\gamma_2 n_B(\omega_\pm) \ll \kappa$, $r \sim {\cal O}(1)$, and $G,|\delta|\ll \omega_{m,2}$, the occupation numbers $n_{\therm,\pm} \ll 1$. In other words, absent nonlinear interaction terms or external driving, the state of the normal modes will be close to vacuum. In experiments to date \cite{Delic2019,Delic2019PRL,Windey2019PRL}, the main contributions to the first term in \eqref{eq:ThermNorm} come from background gas collisions and heating from photon recoil. Here, we will simply assume that these can be made very small.

In principle, there will also be correlations between the normal modes since they couple to common baths. However, as we consider normal mode frequency splitting $\omega_+ - \omega_- \gg \kappa_\pm$, we can ignore this in the following. 

Based on the above considerations, we can think of the normal mode dissipation as if they couple to separate and uncorrelated thermal baths with occupation numbers $n_{\therm,\pm}$. We may then write down an effective quantum master equation for the system density matrix $\rho$:
\begin{align}
\label{eq:Master1}
\dot{\rho} & = -\frac{i}{\hbar} \left[H,\rho \right] \\
& + \sum_{\sigma=\pm}\kappa_\sigma  \left\{ \left(n_{\therm,\sigma} + 1\right) {\cal D}[c_\sigma] +  n_{\therm,\sigma} {\cal D}[c^\dagger_\sigma] \right\} \rho \notag \\
& + \gamma_1 \left\{\left(n_{\therm,1} +1 \right)  {\cal D}[b_1]  +  n_{\therm,1}  {\cal D}[b^\dagger_1] \right\} \rho ,  \notag
\end{align}
with ${\cal D}[o]\rho = o\rho o^\dagger - (o^\dagger o \rho + \rho o^\dagger o)/2 $. While this equation can be used to calculate the system dynamics, we note that it obscures the fact that photons are emitted into the optical bath not only at $\omega_\pm$, but also at $-\omega_\pm$, and that there are correlations between photons emitted at positive and negative frequencies.

\section{Resonant nonlinear interactions}
\label{sec:NonlinInt}

Having established the effective description in terms of normal modes, we proceed to discuss the nonlinear interaction between the normal modes and mechanical mode $b_1$.

\subsection{Effective interaction Hamiltonian}

We now express the Hamiltonian \eqref{eq:Hint1} describing interaction between the cavity mode and the mirror's motion in terms of normal mode operators. This gives 
\begin{align}
\label{eq:HnonlinNormMod}
H_{\mathrm{int},1} & = \hbar x_1  \left[g_{+} c_+^\dagger c_+ +  g_{-} c_-^\dagger c_-  + g_{+-}  \left(c_+^\dagger c_- + c_-^\dagger c_+ \right) \right] \notag \\
& + \hbar \tilde{g}_{+} \left(c_+^{\dagger \, 2} b_1 + b_1^\dagger c_+^2\right) + \hbar \tilde{g}_{-} \left(c_-^{\dagger \, 2} b_1 + b_1^\dagger c_-^2 \right)  \notag \\
& + \hbar \tilde{g}_{+-} \left(c_+^{\dagger} c_-^\dagger b_1 + b_1^\dagger c_- c_+\right)  .
\end{align}
We have neglected terms of the type $c_+^2 b_1$, which will be off resonance by roughly $2\omega_{m,1}$, according to the frequency relation \eqref{eq:RoughFreqRel}. The first line in \eqref{eq:HnonlinNormMod} contains standard radiation pressure interaction terms for both normal modes with $g_+ = g_0/(1+r^2)$ and $g_- = r^2 g_0/(1+r^2)$, as well as cross-terms familiar from two-mode optomechanics \cite{Ludwig2012PRL,Stannigel2012PRL} with $g_{+-} = -r g_0/(1+r^2)$. These terms are all off resonance by $\sim \omega_{m,1}$ with our assumptions, and will not play a significant role.

\begin{center}
\begin{figure}[htb]
\includegraphics[width=.45\textwidth]{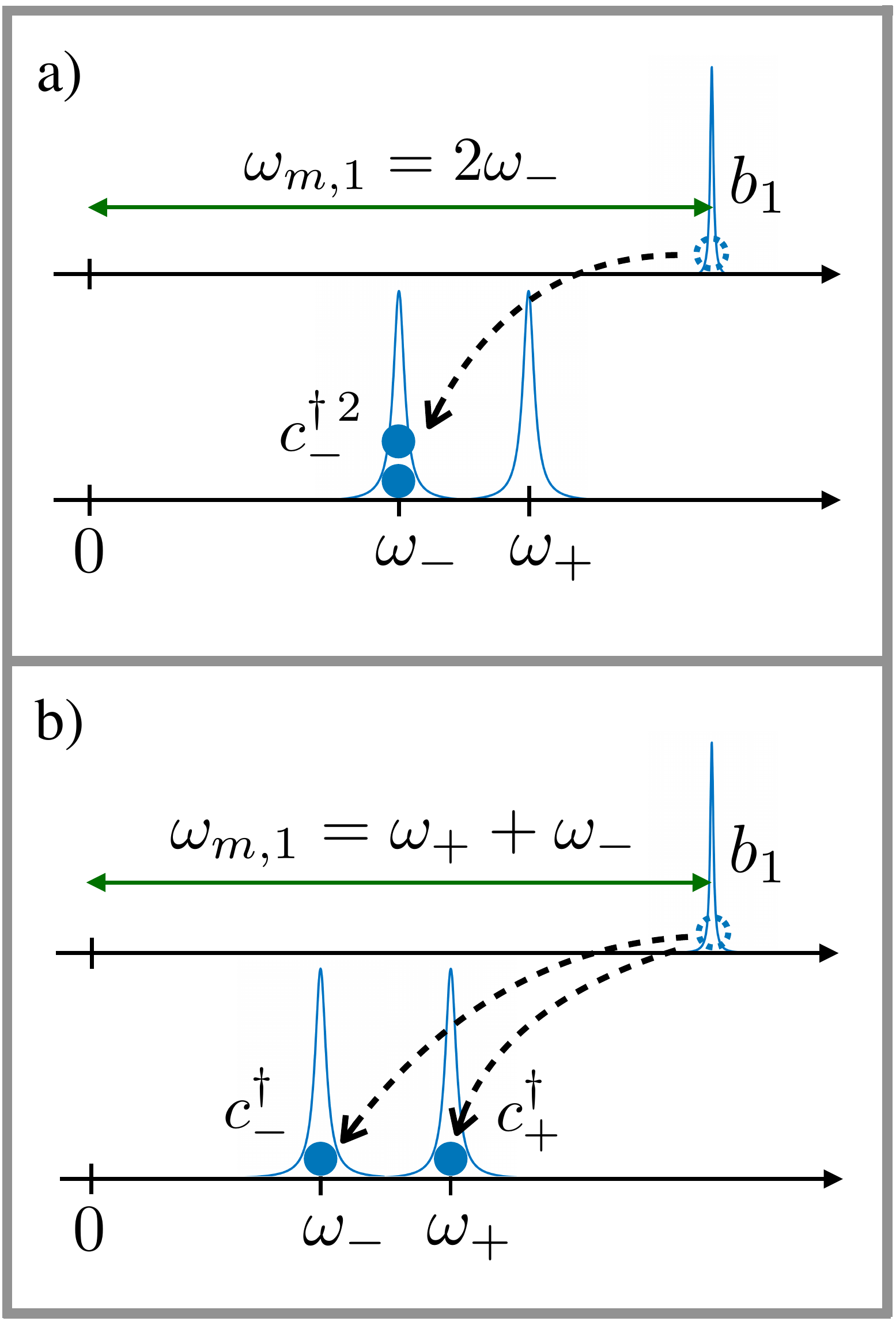}
\caption{Illustration of resonant processes described by the nonlinear optomechanical interaction $H_{\mathrm{int},1}$. a) The process $c_-^{\dagger \, 2} b_1$, where a phonon is annihilated and two particles are created in normal mode $c_-$. b) The process $c_+^{\dagger } c_-^{\dagger } b_1$, where a phonon is annihilated and one particle is created in each normal mode.}
\label{fig:ResNonlinPros}
\end{figure}
\end{center}
In the second line of \eqref{eq:HnonlinNormMod}, we recognize the degenerate parametric oscillator Hamiltonian for both normal modes. These describe processes where two normal mode particles are created and one $b_1$ phonon is destroyed, as illustrated in Figure \ref{fig:ResNonlinPros}a), and vice versa. The effective coupling rates are
\begin{align}
\label{gTilde}
\tilde{g}_{+} & = - \frac{r (G - r\delta/2)}{(1+r^2)^2 \omega_{m,2}} g_0 \\
\tilde{g}_- & = \frac{r^2 (rG + \delta/2)}{(1+r^2)^2 \omega_{m,2} } g_0
\end{align}
and the processes are resonant when 
\begin{equation}
\label{eq:ResCondDPO}
\left| \omega_{m,1} - 2\omega_{\pm} \right| \ll \kappa_\pm .
\end{equation}
For this resonance condition to be satisfied, the frequency of mechanical mode $b_1$ must be in the vicinity of $2\omega_{m,2}$. This is the reason for the requirement \eqref{eq:RoughFreqRel}. However, fine-tuning of the bare frequencies is not a requirement, since \eqref{eq:ResCondDPO} can in principle be met by adjusting the tweezer laser power and detuning. We also emphasize that the resonance condition only needs to be satisfied to well within the normal mode linewidths $\kappa_\pm \sim {\cal O}(\kappa)$.

The third line of \eqref{eq:HnonlinNormMod} is the nondegenerate parametric oscillator Hamiltonian. This describes processes as shown in Figure \ref{fig:ResNonlinPros}b) where one $b_1$ phonon is annihilated and one particle is created in each normal mode, and vice versa. The effective coupling rate is 
\begin{align}
\label{gTildePm}
\tilde{g}_{+-} & = - \frac{r \delta}{2(1+r^2) \omega_{m,2}} g_0
\end{align}
which is nonzero only for detuning $\delta \neq 0$. The processes are resonant if the relations
\begin{equation}
\label{eq:ResCondNDPO}
\left| \omega_{m,1} - (\omega_+ + \omega_-) \right| \ll \kappa_+, \kappa_- 
\end{equation}
are satisfied.

\subsection{Simplified model}

The interaction Hamiltonian \eqref{eq:HnonlinNormMod} can be simplified depending on the choice of frequency relations between the modes.
Let us first consider the frequency relation $|\omega_{m,1} - 2\omega_-| \ll \kappa_-$ such that the nonlinear interaction $\propto \tilde{g}_-$ between normal mode $c_-$ and mechanical mode $b_1$ is resonant. We may then neglect the terms $\propto \tilde{g}_+, \tilde{g}_{+-}$, as their effect will be insignificant compared to those of terms $\propto \tilde{g}_-$ in the regime $G/\kappa \gg 1$. 

In the limit $g_0^2/(\kappa \omega_{m,1}) \ll 1$ that we consider here, the off-resonant radiation pressure terms in the first line of \eqref{eq:HnonlinNormMod} will only become significant if the mechanical mode $b_1$ is excited to very large amplitudes \cite{Braginsky2001PhysLettA,Marquardt2006PRL,Leijssen2017NatComm,Borkje2019PRA}. If we limit ourselves to states with phonon numbers 
\begin{equation}
\label{eq:PhononNumbLimit}
n_1 < \frac{\kappa \omega_{m,1}}{g_0^2} ,
\end{equation}
we can ignore the terms $\propto g_+,g_-, g_{+-}$. We will comment on the validity of this assumption when we consider specific states below. In total, the interaction Hamiltonian reduces to
\begin{align}
\label{eq:HnonlinNormModDPO}
H_{\mathrm{int},1} & =  \hbar \tilde{g}_{-} \left(c_-^{\dagger \, 2} b_1 + b_1^\dagger c_-^2 \right)  
\end{align}
with these assumptions.

Similarly, with the resonance condition \eqref{eq:ResCondNDPO} and the same restriction \eqref{eq:PhononNumbLimit} on mechanical phonon numbers, the effective nonlinear interaction Hamiltonian becomes
\begin{align}
\label{eq:HnonlinNormModNDPO}
H_{\mathrm{int},1} & =  \hbar \tilde{g}_{+-} \left(c_+^{\dagger} c_-^\dagger b_1 + b_1^\dagger c_- c_+\right)  .
\end{align}
i.e., the nondegenerate parametric oscillator model.

\section{Cooling by photon-pair emission}
\label{sec:Cooling}

We now move on to examine the properties of our effective models. In this Section, we start by considering the steady state of the system without any additional driving.

\subsection{Adiabatic elimination}

We consider first the degenerate parametric oscillator Hamiltonian assuming the frequency relation $\omega_{m,1} - 2\omega_- \ll \kappa_-$. The choice of resonance with mode $c_-$ is arbitrary -- we could just as well have chosen the other normal mode $c_+$. 

In the regime $\gamma_1 n_{\therm,1}  , \tilde{g}_- \ll \kappa_-$ that we consider, the state of the normal mode $c_-$ will be largely unaffected by the interaction as long as the system is not driven. After moving to rotating frames such that $c_- \rightarrow e^{-i\omega_{m,1}t/2} c_-$ and $b_1\rightarrow e^{-i \omega_{m,1} t} b_1$, adiabatic elimination \cite{Breuer2007Book} of the mode $c_-$ from Equation \eqref{eq:Master1} gives an effective master equation for the reduced density matrix $\rho_m$ of the mechanical mode:
\begin{align}
\label{eq:Masterrhom}
\dot{\rho}_m & =  \tilde{\gamma}_1 \left\{\left(\tilde{n}_{\therm,1} +1 \right)  {\cal D}[b_1]  +  \tilde{n}_{\therm,1}  {\cal D}[b^\dagger_1] \right\} \rho_m .  \notag
\end{align}
From this, we can conclude that the mechanical mode $b_1$ is in a thermal state with an average occupation number
\begin{equation}
\label{eq:AvgPhononNumb}
\tilde{n}_{\therm,1} = \langle b_1^\dagger b_1 \rangle = \frac{\gamma_1 n_\mathrm{th,1}}{\tilde{\gamma}_1} ,
\end{equation}
where the effective mechanical linewidth is defined as
\begin{equation}
\label{eq:EffMechLineWDPO}
\tilde{\gamma}_{1} = \gamma_1 + \frac{4 \tilde{g}_-^2}{\kappa_-} \left( 1 + 2n_{\therm,-} \right) .
\end{equation}
We have neglected terms $\propto n_{\therm,-}^2 \ll 1$, which is a good approximation as long as $n_{\therm,-}^2 \ll \tilde{n}_{\therm,1}$. 

The physical interpretation of Equations \eqref{eq:AvgPhononNumb} and \eqref{eq:EffMechLineWDPO} is that the mechanical mode is cooled as a result of the nonlinear interaction. The additional decay rate 
\begin{equation}
\label{eq:SpontDecRate}
\Gamma_1 = \frac{4 \tilde{g}_-^2}{\kappa_-} ,
\end{equation} 
appearing in \eqref{eq:EffMechLineWDPO} is due to {\it spontaneous} emission of two normal mode particles that subsequently decay, primarily as photons into the optical bath. The last term $2 \Gamma_1 n_{\therm,-}$ results from additional {\it stimulated} emission caused by the (small) thermal occupation of the normal mode $c_-$.

For completeness, let us also consider the nondegenerate case \eqref{eq:HnonlinNormModNDPO} with the resonance condition \eqref{eq:ResCondNDPO}. Adiabatic elimination of the normal modes again leads to a thermal mechanical state with average occuation number \eqref{eq:AvgPhononNumb}. However, the effective mechanical linewidth now becomes
\begin{equation}
\label{eq:EffMechLineWNDPO}
\tilde{\gamma}_{1} = \gamma_1 + \frac{4 \tilde{g}_{+-}^2}{\kappa_+ + \kappa_-} \left( 1 + n_{\therm,+} + n_{\therm,-} \right) .
\end{equation}
The interpretation is the same -- the mechanical mode is cooled by processes where one phonon is converted to two normal mode particles that subsequently decay. The difference is that the two normal mode particles now enter separate modes at separate frequencies.

The validity of our simplified models relies on the assumption \eqref{eq:PhononNumbLimit}. We note that this is satisfied as long as $\tilde{n}_{\mathrm{th},1} \ll \kappa \omega_{m,1}/g_0^2$. The latter is a very large number in the weak-coupling regime, such that this can be fulfilled even for significant thermal occupation of the mechanical baths.

\subsection{Effective cooperativity}
\label{sec:EffCoop}

In the following, we will again focus on the degenerate parametric oscillator interaction \eqref{eq:HnonlinNormModDPO} with $\omega_{m,1} \approx 2\omega_-$. We now define an effective cooperativity ${\cal C}_-$ as the ratio between the decay rate $\Gamma_1$ due to the above-mentioned spontaneous two-photon emission and the intrinsic mechanical decay rate $\gamma_1$, giving
\begin{equation}
\label{eq:EffCoop}
{\cal C}_{-} = \frac{\Gamma_1}{\gamma_1} =  \frac{4 \tilde{g}_-^2}{\kappa_- \gamma_1} .
\end{equation}
Note that $\tilde{n}_{\therm,1} \approx n_{\therm,1}/{\cal C}_-$ for $n_{\therm,-} \ll 1$. To relate to previous work on the degenerate parametric oscillator, ${\cal C}_- \gtrsim 1$ corresponds to the so-called finite-system size parameter regime for the fundamental mode \cite{CarmichaelBook2008,Veits1995PRA,Veits1997PRA,Veits1997PRA_2}.

For the simplified case of detuning $\delta = 0$, i.e., $r = 1$, we have $\tilde{g}_- = g_0 G/(4\omega_{m,2})$ and $\kappa_- \approx \kappa/2$, which gives 
\begin{equation}
\label{eq:EffCoop2}
{\cal C}_{-}  = 2  \left(\frac{G}{4\omega_{m,2}}\right)^2 {\cal C}_0 .
\end{equation}
Here, ${\cal C}_0$ is the single-photon optomechanical cooperativity defined in Equation \eqref{eq:SinglePhotCoopDef}. We have already assumed the parameter hierarchy $\kappa < G < \omega_{m,2}$, so a natural choice for $G$ would be the geometric mean of $\kappa$ and $\omega_{m,2}$. If we let $G =  (8 p \kappa \omega_{m,2})^{1/2}$ with $p$ a dimensionless number of order unity, we get
\begin{equation}
\label{eq:EffCoop3}
{\cal C}_- = p \frac{\kappa}{\omega_{m,2}} {\cal C}_0 .
\end{equation}
From this, we see that the effective cooperativity ${\cal C}_-$ can exceed unity, but only if the single-photon cooperativity is large, i.e., ${\cal C}_0 > \omega_{m,2}/\kappa \gg 1$. 

Let us now contrast this with what would happen in a different model with direct driving of an optical cavity dispersively coupled to two mechanical oscillators. One would then have an additional term $H' = \hbar G x_1 (a + a^\dagger)$ in the original Hamiltonian, which would renormalize the mechanical decay rate $\gamma_1 \rightarrow \gamma_1 +  G^2 \kappa/(2 q \, \omega_{m,2}^2)$ due to up- and down-conversion of drive photons, with $q$ a dimensionless number of order unity \cite{Borkje2013PRL}. Equation \eqref{eq:EffCoop2} would then give ${\cal C}_- < q (g_0/\kappa)^2 \ll 1$ in the limit of weak single-photon coupling. This shows that if we want to realize a large effective cooperativity associated with the nonlinear and resonant interaction processes, it is essential to not have any direct driving of the cavity mode.

\section{Second harmonic generation}
\label{sec:SHD}

In this Section, we analyze the behaviour of our system when subjected to an optical probe. We again focus on the degenerate parametric oscillator model \eqref{eq:HnonlinNormModDPO} with $\omega_{m,1} \approx 2\omega_-$, but note that the results presented can be straightforwardly generalized to the nondegenerate model as well.

\begin{center}
\begin{figure}[htb]
\includegraphics[width=.4\textwidth]{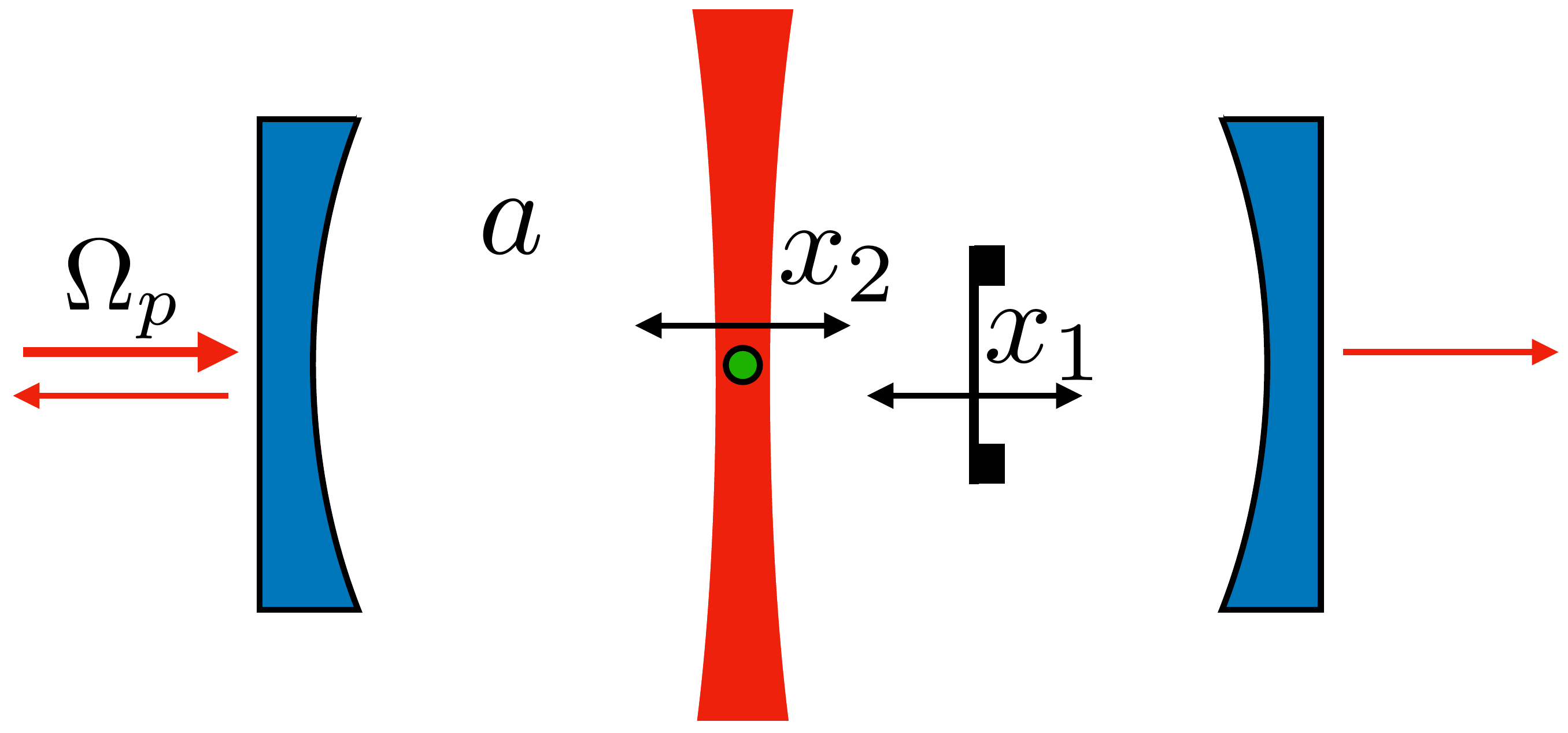}
\caption{Schematic of a modified setup with a membrane-in-the-middle rather than a movable end mirror. The reponse to an optical probe can be measured either in transmission or reflection.}
\label{fig:Probe}
\end{figure}
\end{center}

We will now consider a setup with a two-sided cavity such that light can be detected in transmission. This is for example possible if, rather than a movable end mirror as in Figure \ref{fig:CSModel}, mode $b_1$ is a flexural mode of a dielectric membrane inside the cavity. Such a modified setup is illustrated in Figure \ref{fig:Probe}. 

\subsection{Optical probe}

To probe the system, we add a coherent optical drive at a frequency $\omega_p = \omega_\mathrm{tw} + \omega_{m,1}/2 + \Delta_p$, where we define $\Delta_p$ as the probe detuning. In the frame rotating at $\omega_\mathrm{tw}$, in which \eqref{eq:Hfree}, \eqref{eq:Hint1}, and \eqref{eq:Hint2} are expressed, this corresponds to adding a term 
\begin{equation}
\label{eq:Hprobe}
H_\mathrm{probe} = i \hbar \Omega_p \left(e^{i (\omega_{m,1}/2 + \Delta_p)t} a - e^{-i (\omega_{m,1}/2 + \Delta_p)t} a^\dagger \right) .
\end{equation}
to the Hamiltonian. Expressing this in terms of the operators $c_\pm$ gives rise to drive terms for both normal modes. However, the probe will be off resonance with the normal mode frequency $\omega_+$, such that roughly half of the probe power will be promptly reflected from the cavity. This is not relevant to the response of mode $c_-$, but it will influence the overall transmitted or reflected probe power.

We now write down the quantum Langevin equations corresponding to the effective master equation \eqref{eq:Master1} with the addition of the coherent probe. In the frame rotating at $\omega_p$, i.e., $c_- \rightarrow e^{-i(\omega_{m,1}/2 + \Delta_p) t} c_-$ and $b_1 \rightarrow e^{-i(\omega_{m,1} + 2\Delta_p) t} b_1$, they become
\begin{align}
\label{eq:QLEff1} \dot{c}_- & = - \frac{\kappa_-}{2} c_- - 2 i \tilde{g}_- c_-^\dagger b_1 + \Omega_- + \sqrt{\kappa_-} \zeta_- , \\
\label{eq:QLEff2} \dot{b}_1 & = -  \left(\frac{\gamma_1}{2} - 2 i \Delta_p \right) b_1 - i \tilde{g}_- c_-^2 + \sqrt{\gamma_1} \eta_1 , 
\end{align}
when we, for simplicity, assume $|\Delta_p| \ll \kappa_-$ and define
\begin{equation}
\label{eq:OmegaMDef}
\Omega_- = \frac{r\left[1+\frac{1}{2}(q_-^2 + p^2)\right]}{\sqrt{1+r^2}} \Omega_p  .
\end{equation}
We have also defined standard Gaussian white noise operators that satisfy the commutation relations
\begin{equation}
\label{eq:CommRel}
\left[\zeta_-(t) , \zeta_-^\dagger(t') \right] = \left[\eta_1(t) , \eta_1^\dagger(t') \right] = \delta(t-t') 
\end{equation}
and have the properties
\begin{align}
\label{eq:NoiseProp}
\langle \zeta_-^\dagger(t) \zeta_-(t') \rangle = n_{\therm,-} \, \delta(t-t')  , \\
\langle \eta_1^\dagger(t) \eta_1(t') \rangle = n_{\therm,1} \, \delta(t-t') .
\end{align}
Note that $\zeta$ is defined so as to comply with the effective master equation \eqref{eq:Master1} and can thus not be directly read out from the original quantum Langevin equation \eqref{eq:QLM}. 

\subsection{Classical approximation}

The optical probe will not only lead to a nonzero normal mode coherence $\langle c_- \rangle$, but will also cause coherent mechanical oscillations such that $\langle b_1 \rangle \neq 0$. We argued in Section \ref{sec:Cooling} that the thermal motion of the oscillator alone will not affect the mode $c_-$ significantly. It is only when driven to coherent amplitudes much larger than the thermal motion that the oscillator can begin to influence the mode $c_-$.  As a first approximation, it thus seems reasonable to simply replace the operator $b_1$ with a complex number, i.e., $b_1 \rightarrow -i\beta$.

The most naive thing we can do is to also replace $c_-$ by a complex number, $c_- \rightarrow \alpha_-$, thereby ignoring quantum and thermal fluctuations in mode $c_-$. In the steady state and with these approximations, Equations \eqref{eq:QLEff2} gives
\begin{align}
\label{eq:QLEffSol2} \beta =    \frac{\tilde{g}_-}{\gamma_1/2 - 2i \Delta_p} \alpha_-^2  , 
\end{align}
and the normal mode amplitude is determined by the third-order equation
\begin{align}
\label{eq:QLEffSol1} \alpha_- & = \frac{\alpha_{-}^{(0)}}{1 + 2 {\cal C}_- |\alpha_-|^2 /(1 - 4i \Delta_p/\gamma_1) }  ,
\end{align}
according to Equation \eqref{eq:QLEff1}. Here, we have defined 
\begin{equation}
\label{eq:alphaM0}
\alpha_{-}^{(0)} = \frac{2 \Omega_-}{\kappa_-} .
\end{equation}
which is the coherent amplitude of normal mode $c_-$ in absence of interactions. 

Equation \eqref{eq:QLEffSol1} shows that the coherent response of normal mode $c_-$ is nonlinear in probe power within a narrow frequency interval. This effect, which we analyze further below, can be viewed as a nonlinear version of optomehanically induced transparency \cite{Agarwal2010PRA,Weis2010Science}, i.e., a suppression of coherence due to destructive interference. A similar phenomenon was studied theoretically for strongly driven single-mode optomechanics \cite{Borkje2013PRL,Lemonde2013PRL,Kronwald2013PRL}. We note that for small ${\cal C}_-|\alpha_-|^2$, Equation \eqref{eq:QLEffSol1} reproduces Equation (12) in Ref.~\cite{Borkje2013PRL}. 

Let us for a moment consider the case when the probe is on resonance, i.e., $\Delta_p = 0$. It is then well known \cite{McNeil1978OptComm,Drummond1980OpticaActa} that the steady state solutions \eqref{eq:QLEffSol2} and \eqref{eq:QLEffSol1} become unstable at a critical probe drive, which in the limit $\kappa_- \gg \gamma_1$ corresponds to 
\begin{align}
\label{eq:CritDrive}
\alpha_{-}^{(0)} =  \alpha_{-,\text{crit}}^{(0)} \equiv  \sqrt{\frac{2}{{\cal C}_-}}  .
\end{align}
At this drive strength, the mechanical oscillation amplitude reaches the critical value
\begin{align}
\label{eq:Critbeta} \beta = \beta_\text{crit} =   \frac{\kappa_-}{4\tilde{g}_-}  , 
\end{align} 
again assuming $\kappa_- \gg \gamma_1$. We note that $\beta_\text{crit} \gg 1$ with our assumptions. For probe strengths beyond the critical value, the system can settle into a limit cycle characterized by self-pulsing of the coherent amplitudes \cite{McNeil1978OptComm,Drummond1980OpticaActa}. However, in the limit $\kappa_- \gg \gamma_1$, the amplitudes of the limit-cycle oscillations decrease and eventually become smaller than the size of thermal or vacuum fluctuations \cite{Veits1997PRA}.  

The above approximation fails to account for the fact that the mechanical oscillator is damped due to spontaneous conversion of phonons to pairs of normal mode particles, as discussed in Section \ref{sec:Cooling}. Naively, one would think that this approximation becomes invalid for sufficiently small intrinsic mechanical decay rates $\gamma_1$ such that the effective cooperativity ${\cal C}_- \sim 1$. However, we will see below that, due to the coherent mechanical oscillations, the additional damping channel eventually becomes relevant even for arbitrarily small ${\cal C}_-$.

Earlier, we restricted our model to phonon numbers such that \eqref{eq:PhononNumbLimit} is valid. At the critical mechanical amplitude \eqref{eq:Critbeta}, we have 
\begin{align}
\label{eq:CritbetaOrigPar} \frac{g_0^2}{\kappa \omega_{m,1}} |\beta_\text{crit}|^2 \sim \frac{\kappa \omega_{m,1}}{G^2}  .
\end{align} 
This means that the assumption \eqref{eq:PhononNumbLimit} is valid for $\beta \lesssim \beta_\text{crit}$ as long as $G \gtrsim \sqrt{\kappa \omega_{m,1}}$, which fits well with our assumption of a parameter regime $\kappa < G < \omega_{m,1}$.

\subsection{Semiclassical approximation}
\label{sec:SC}
We will now include fluctuations in mode $c_-$. This means that we retain the operator $\zeta_-$ in Equation \eqref{eq:QLEff1}. However, we still ignore fluctuations in mode $b_1$, replacing the operator $b_1 \rightarrow -i\beta$ with a constant complex number as before. Equation \eqref{eq:QLEff1} then becomes the standard equation of motion for a degenerate parametric amplifier, but one where the squeezing parameter $\beta$ actually depends on the probe drive and must be determined self-consistently. In other words, the depletion of the harmonic amplitude $\beta$ due to decay back to the fundamental mode $c_-$ must be taken into account, as was done in Refs.~\cite{Veits1997PRA,Veits1997PRA_2}.

It is convenient to rescale the mechanical amplitude to the critical value through the definition
\begin{equation}
\label{eq:BDef}
B =\frac{\beta}{ \beta_\text{crit} } .
\end{equation}
Solving Equation \eqref{eq:QLEff1} then gives the normal mode coherence
\begin{equation}
\label{eq:alphaDef}
\alpha_- = \langle  c_- \rangle = \frac{1 - B}{1 - |B|^2} \alpha_-^{(0)} 
\end{equation}
and the average occupation number
\begin{equation}
\label{eq:nDef}
n_- = \langle  c_-^\dagger c_- \rangle = |\alpha_-|^2 + \frac{n_{\therm,-} + |B|^2/2}{1 - |B|^2}  .
\end{equation}
We also calculate the expectation value
\begin{equation}
\label{eq:sigmaDef}
\sigma_- = \langle  c_-^2 \rangle = \alpha_-^2 - \frac{n_{\therm,-} + 1/2}{1 - |B|^2}B .
\end{equation}
By taking the expectation value of Equation \eqref{eq:QLEff2} and inserting \eqref{eq:sigmaDef}, we get the following self-consistency equation for the coherent mechanical amplitude:
\begin{equation}
\label{eq:EOMB}
\left(1 -  \frac{4i \Delta_p}{\gamma_1}+  \frac{{\cal C}_- \left(1 + 2 n_{\therm,-}\right)}{1 - |B|^2}\right)B = \frac{4 (1 - B)^2 \tilde{\Omega}_-^2}{\left(1 - |B|^2\right)^2} .
\end{equation}
Here, we have rescaled the drive parameter
\begin{equation}
\label{eq:OmegaTildeDef}
\tilde{\Omega}_- = \sqrt{\frac{{\cal C}_-}{2}}  \alpha_-^{(0)} = \sqrt{2 {\cal C}_-} \, \frac{\Omega_-}{\kappa_-}
\end{equation}
such that $\tilde{\Omega}_- = 1$ at the critical drive strength. The first term in the paranthesis on the left hand side of \eqref{eq:EOMB} originates from the intrinsic mechanical damping, whereas the last term in the paranthesis describes additional nonlinear damping due to emission of incoherent normal mode particles. We see that for mechanical amplitudes $|B| \ll 1$, the latter is relevant when the cooperativity ${\cal C}_-$ becomes comparable to one, as expected. In general, it is relevant when $1 - |B|^2 \lesssim {\cal C}_-$.
\begin{center}
\begin{figure}[htb]
\includegraphics[width=.45\textwidth]{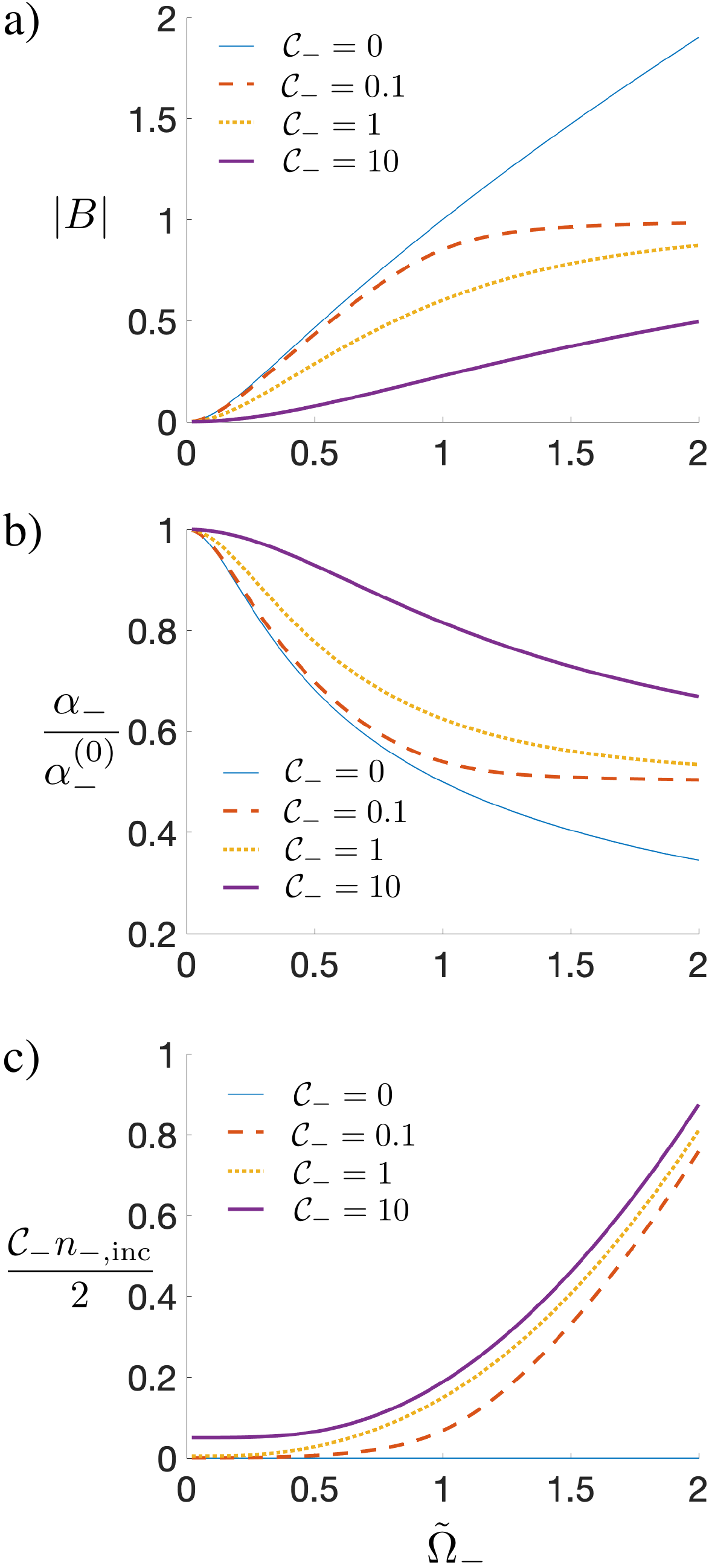}
\caption{a) Coherent response of the mechanical oscillator vs.~optical probe drive strength for $\Delta_p = 0$. b) Coherent response of normal mode $c_-$ for $\Delta_p = 0$ vs.~optical probe drive strength normalized to the coherence in absence of interactions. c) Normalized average occupation number of incoherent normal mode particles vs.~drive strength for $\Delta_p = 0$. We have assumed $n_{\therm,-} = 0.01$. }
\label{fig:SemiClass}
\end{figure}
\end{center} 

We display the numerical solution to the self-consistency Equation \eqref{eq:EOMB} as a function of probe drive strength in Figure \ref{fig:SemiClass}a). We have limited ourselves to $|B| < 1$ and $\Delta_p = 0$, in which case one can show that Equation \eqref{eq:EOMB} has a fixed point such that $B = |B|$. For small cooperativity ${\cal C}_-$, we see that $|B|$ follows the classical solution \eqref{eq:QLEffSol2} for small drive strengths, but deviates from it when approaching the critical drive strength $\tilde{\Omega}_- = 1$. For larger cooperativities, the deviation from the classical solution is more pronounced and evident also for small drive strengths. 

For $\Delta_p = 0$ and $B = |B|$ real, we can find an analytic solution to Equation \eqref{eq:EOMB} in the regime ${\cal C}_- \gg 1 - |B|^2$. This can be written
\begin{equation}
\label{eq:SolAnalyticB}
|B| = \frac{1}{2}\left(\sqrt{(1 + \lambda)^2 + 4\lambda} - (1 + \lambda) \right)  
\end{equation}
when defining $\lambda = 4 \tilde{\Omega}_-^2/({\cal C}_- (1+2n_{\therm,-}))$. Taking the limit $\lambda \rightarrow \infty$ gives $|B| \rightarrow 1$, reproducing the behavior shown in Figure \ref{fig:SemiClass}a). 

We see that for any nonzero ${\cal C}_-$, the mechanical coherence will never reach the critical value $|B| = 1$. However, the approximation $b_1 \rightarrow - i \beta$ will break down as $|B|$ gets sufficiently close to 1, at which point mechanical fluctuations must also be taken into account \cite{Veits1997PRA,Veits1997PRA_2}. We will comment further on this in Section \ref{sec:Numerics}.

\begin{center}
\begin{figure}[htb]
\includegraphics[width=.45\textwidth]{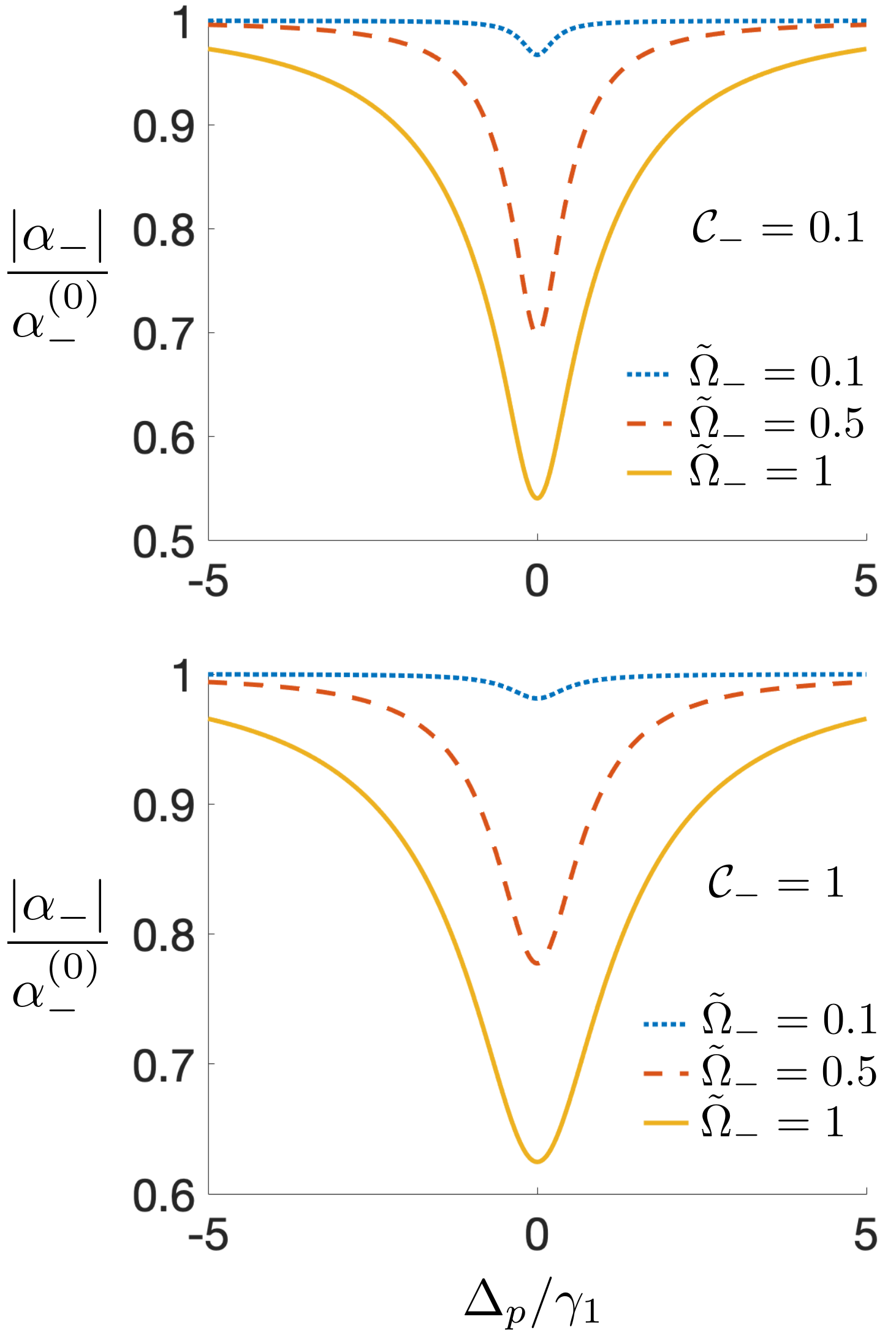}
\caption{ Coherent response of normal mode $c_-$ vs.~probe detuning $\Delta_p$, shown for different optical probe drive strengths and normalized to the coherence in absence of interactions. This can be viewed as a nonlinear version of optomechanically induced transparency. In this case, both the depth and width of the transparency dip depends on probe power. Note that for ${\cal C}_- \gtrsim 1$, the response is strongly nonlinear in probe power even in the few-photon regime $|\alpha_-| \lesssim 1$. }
\label{fig:OMITFreq}
\end{figure}
\end{center}
In Figure \ref{fig:OMITFreq}, we plot the normal mode coherent amplitude $|\alpha_-|$ as a function of detuning $\Delta_p$ for different values of cooperativity ${\cal C}_-$ and drive strengths $\tilde{\Omega}_-$. This shows a dip around $\Delta_p = 0$ due to destructive interference, whose width is given by the effective mechanical linewidth. This is the nonlinear analog of optomechanically induced transparency \cite{Borkje2013PRL,Lemonde2013PRL,Kronwald2013PRL} referred to above. We observe that both the size and the width of the dip depends on drive strength $\tilde{\Omega}_-$. In Figure \ref{fig:SemiClass}b), we display the relative suppression of coherence $\alpha_-/\alpha_{-}^{(0)}$ at the bottom of the dip, i.e., for $\Delta_p = 0$, as a function of drive strength. We note that as $\tilde{\Omega}_- \rightarrow \infty $ and $B \rightarrow 1$, Equation \eqref{eq:alphaDef} gives $\alpha_-/\alpha_{-}^{(0)} \rightarrow 1/2$. 

The interference dip shown in Figure \ref{fig:OMITFreq} can occur even for cooperativites ${\cal C}_- < 1$ when compensating with stronger probe drives. However, it is worth noting that, in the regime ${\cal C}_- \gtrsim 1$, the normal mode has a significant nonlinear response already in the few-particle regime. This is clear from noting that when $\tilde{\Omega}_- = 1$, we have $|\alpha_-^{(0)}|^2 = |\alpha_{-,\mathrm{crit}}^{(0)}|^2  =  2/{\cal C}_-$. 

The nonlinear mechanical damping due to quantum and thermal fluctuations leads to pairs of incoherent normal mode particles. In Figure \ref{fig:SemiClass}c), we show the incoherent contribution to the average occupation number $n_-$, defined as
\begin{equation}
\label{eq:nincDef}
n_{-,\text{inc}} = n_- - |\alpha_-|^2 
\end{equation}
normalized to $|\alpha_{-,\mathrm{crit}}^{(0)}|^2 = 2/{\cal C}_-$. Again, we see that the relative deviations from the classical result grows for increasing ${\cal C}_-$. The critical behaviour at $\tilde{\Omega}_- = 1$ predicted by classical theory is smeared out, mainly due to quantum fluctuations.

\subsection{Photon antibunching}
\label{sec:Antibunching}

The degenerate parametric amplifier is known to exhibit squeezing, meaning that fluctuations in one quadrature of the mode $c_-$ will be suppressed at the expense of amplified fluctuations in the conjugate quadrature. The unusual feature in our case is that the phase and amplitude of the squeezing parameter are not externally controlled parameters, but they are dependent on the probe drive itself. This has the useful consequence that the normal mode will always be amplitude squeezed, as long as $|\Delta_p|/\gamma_1 \ll \text{max} (1, {\cal C}_-)$, irrespective of the phase of the probe drive. Furthermore, for large cooperativity ${\cal C}_-$, the normal mode will feature significant amplitude squeezing for a wide range of drive strengths in the few-photon regime $|\alpha_-| \lesssim 1$. The system can then display the nonclassical phenomenon of antibunching due to destructive two-photon interference \cite{Stoler1974PRL,Walls1983Nature,Mahran1986PRA,Lu2002PRL,Lemonde2014PRA}.

We define the second order coherence at zero time as
\begin{equation}
\label{eq:g2Def}
g_{-}^{(2)} = \frac{\langle c_-^{\dagger \, 2} c_-^2 \rangle }{ \langle c_-^\dagger c_- \rangle^2 }
\end{equation}
for the normal mode $c_-$, which can be measured through coincidence counting on the photons exiting the cavity in transmission. In the approximation $b_1 \rightarrow -i \beta$, the normal mode $c_-$ is in a Gaussian state, such that higher-order correlation functions can be expressed as second order correlation functions. Assuming $\Delta_p = 0$, this gives
\begin{equation}
\label{eq:g2Gauss}
g_{-}^{(2)} = 1 + \frac{2\alpha_-^2 \left(n_{-,\mathrm{inc}} + \sigma_{-,\mathrm{inc}}\right) + n_{-,\mathrm{inc}}^2 + \sigma_{-,\mathrm{inc}}^2}{\left(\alpha_-^2 + n_{-,\mathrm{inc}} \right)^2}
\end{equation}
where $\alpha_-$ is real and we have defined $\sigma_{-,\mathrm{inc}} = \sigma_- - \alpha_-^2$. We note that $\sigma_{-,\mathrm{inc}}$ is negative, which means that antibunching, i.e., $g_{-}^{(2)} < 1$, is possible for a particular set of parameters. 

\begin{center}
\begin{figure}[htb]
\includegraphics[width=.45\textwidth]{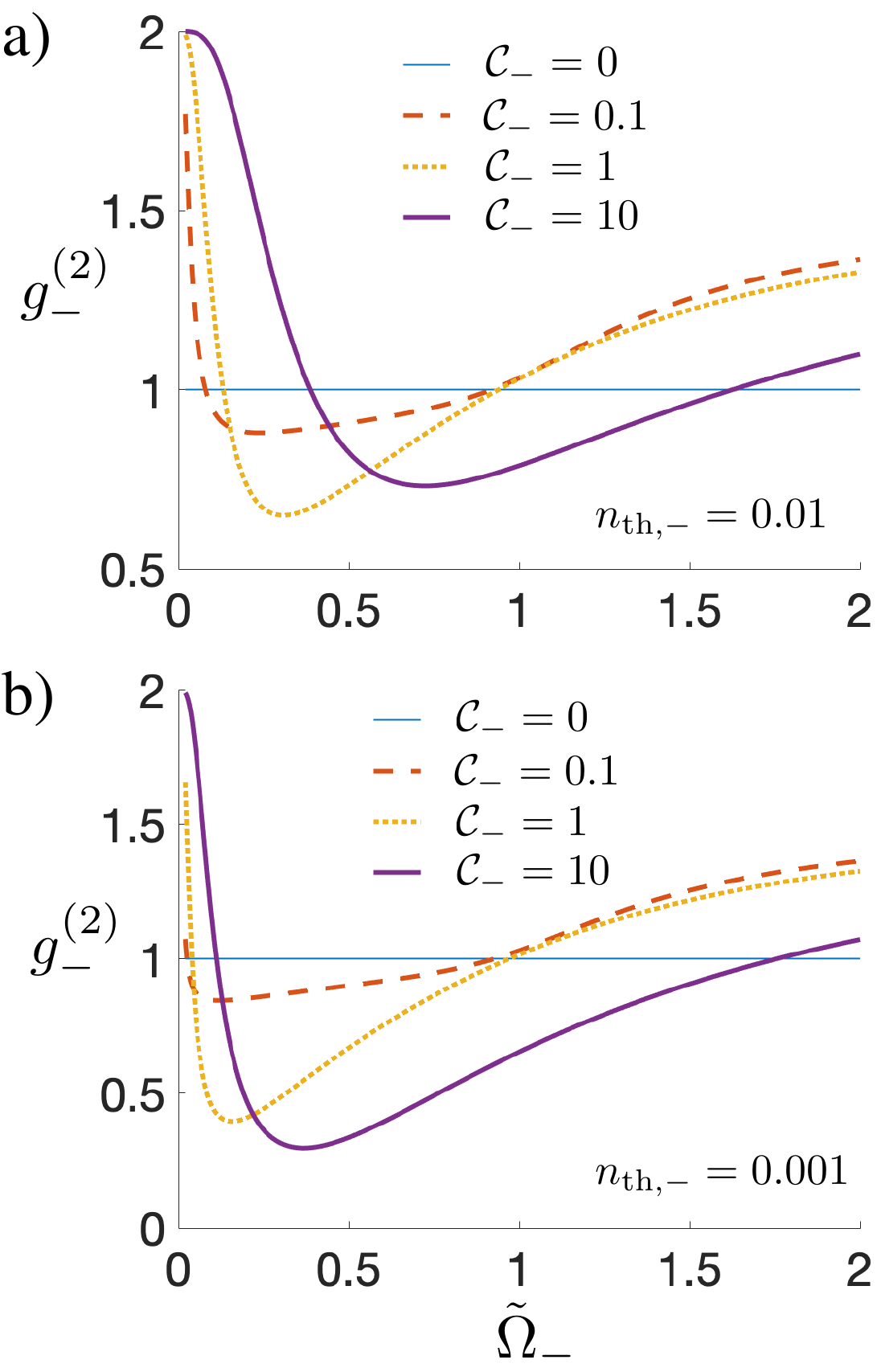}
\caption{The second order coherence at zero time vs.~optical probe drive strength. The normal mode $c_-$ is automatically amplitude squeezed. For ${\cal C}_- \gtrsim 1$, this occurs over a wide range of drive strengths in the few-photon regime $\alpha_- \lesssim 1$, giving rise to antibunching.}
\label{fig:g2M}
\end{figure}
\end{center} 

Figure \ref{fig:g2M} shows the equal time second order coherence $g_{-}^{(2)}$ as a function of drive strength for different values of cooperativity ${\cal C}_-$ and for different values of thermal occupation $n_{\therm,-}$ in the normal mode. We observe significant antibunching for cooperativities ${\cal C}_- \gtrsim 1$. 

For mechanical amplitudes $|B| \ll 1$ and in the limits ${\cal C}_- \gg 1$, $n_{\therm,-} \rightarrow 0$, Equations \eqref{eq:EOMB} and \eqref{eq:alphaDef} give $|B| \approx 2\alpha_-^2$. This corresponds to the optimal amount of squeezing required for minimizing $g^{(2)}_-$ in the limit of small occupation $\alpha_-^2 \ll 1$ and for $n_{\therm,-} \rightarrow 0$ \cite{Lemonde2014PRA}. In our case, the optimal mechanical amplitude self-adjusts to the probe drive without the need for independent tuning of an external squeezing parameter. From Figure \ref{fig:g2M}, we observe that for nonzero $n_{\therm,-}$, the optimal squeezing is not necessarily reached by maximizing ${\cal C}_-$.

Not surprisingly, the antibunching effect vanishes as the thermal occupation number $n_{\therm,-}$ increases \cite{Lemonde2014PRA}. For this reason, one would ideally want to go the limit $n_{\therm,-} \rightarrow 0$. However, we must remember that both the resonant nonlinear interaction and the quantum heating contribution to the thermal occupation originate from the counter-rotating terms in the original Hamiltonian \eqref{eq:Hint2}. If we for example let $r =1$, we get $n_{\therm,-} \geq G^2/(4\omega_{m,2}^2)$, which again gives
\begin{equation}
\label{eq:CoopLimit}
{\cal C}_- <  \frac{{\cal C}_0 n_{\therm,-}}{2} .
\end{equation}
This shows that the limit $n_{\therm,-} \rightarrow 0$ requires the single-photon cooperativity $ {\cal C}_0 \rightarrow \infty$ in order to reach ${\cal C}_- \sim  1$.

\subsection{The nonlinear regime}
\label{sec:Numerics}

We have so far neglected fluctuations around the large coherent mechanical amplitude, in which case the state of the normal mode is a Gaussian, displaced squeezed state. This leads to large fluctuations in the anti-squeezed phase quadrature which decay at a rate $\kappa_- (1-|B|)$. For large enough probe powers, the mechanical amplitude $|B|$ will get sufficiently close to 1 such that this decay rate is comparable to the effective coupling rate $\tilde{g}_-$. At this point, the approximation $b_1 \rightarrow -i\beta$ breaks down and mechanical fluctuations must be taken into account. One then enters a regime of fully nonlinear dynamics and non-Gaussian steady states.

The full nonlinear dynamics is not straightforward to describe analytically. There are some known analytical results for quasiprobability distributions in the limit of a fast decaying harmonic mode \cite{CarmichaelBook2008}, but we consider the opposite limit here. Other approaches rely on perturbative expansions around the semiclassical solution in Section \ref{sec:SC} using non-equilibrium Green's function techniques \cite{Veits1997PRA,Veits1997PRA_2}.

We have solved the quantum master equation numerically using the software QuTiP \cite{Johansson2012CPC,Johansson2013CPC} in order to find the steady state in presence of the probe drive. This involved truncating Hilbert space to a finite number of Fock states, after having displaced both modes by a suitable complex amplitude. We define $W_c(\alpha_-)$ as the Wigner quasiprobability distribution \cite{Walls2008Book,CarmichaelBook2008} of the reduced density matrix of the normal mode $c_-$, and $W_m(\beta)$ as the Wigner distribution of the reduced density matrix of the mechanical mode $b_1$.

\begin{center}
\begin{figure}[htb]
\includegraphics[width=.45\textwidth]{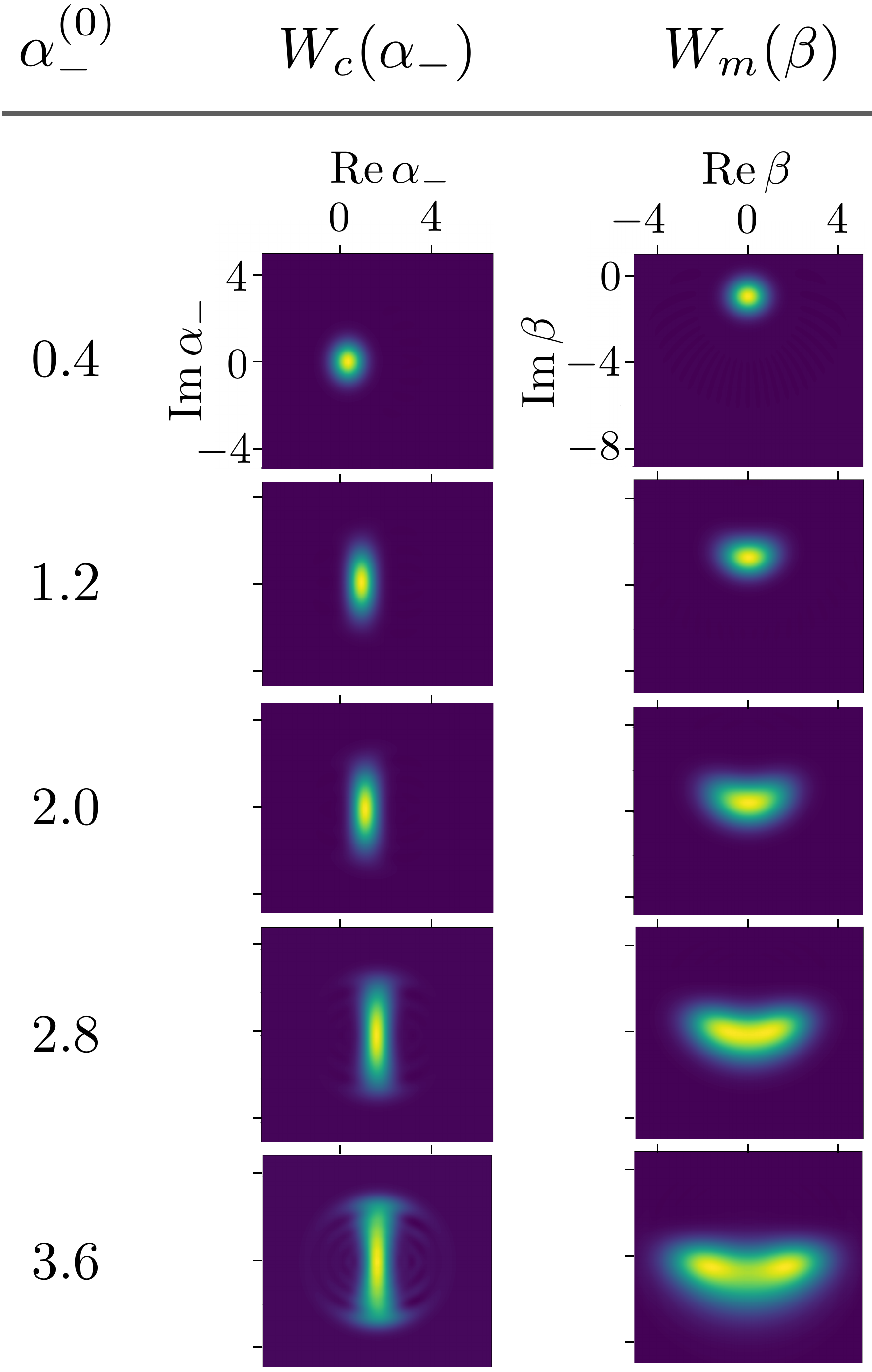}
\caption{Numerical results for the Wigner quasiprobability distributions $W_c(\alpha_-)$ and $W_m(\beta)$ of the normal mode $c_-$ and mechanical mode $b_1$, respectively, for different values of the probe drive strength parametrized by $\alpha_-^{(0)}$. Parameters are chosen such that $\beta_\mathrm{crit} = 4$, ${\cal C}_- = 15.6$, $n_{\therm,-} = 0.01$ and $n_{\therm,1} = 1$.}
\label{fig:Numerics}
\end{figure}
\end{center}

In Figure \ref{fig:Numerics}, we present the Wigner distributions for different values of the probe drive strength, assuming a critical amplitude $\beta_\mathrm{crit} = 4$ and an effective cooperativity ${\cal C}_- = 15.6$. Ideally, we would like to study the regime of $\beta_\mathrm{crit} = \kappa_-/(4\tilde{g}_- ) \gg 1$, but this requires more computational resources than we have applied. We have set the thermal occupation numbers to $n_{\therm,-} = 0.01$ and $n_{\therm,1} = 1$. The latter choice is again due to the necessary Hilbert space truncation.

Figure \ref{fig:Numerics} shows that for small drive strengths, the mechanical state is a displaced thermal state and the normal mode state is an amplitude-squeezed state. This is as expected from the semiclassical analysis. For larger drive strengths, when the average mechanical amplitude gets close to $\beta_\mathrm{crit} = 4$, we see that the mechanical phase fluctuations grows. The reason is that the coherent force on the mechanical oscillator from the normal mode has large phase fluctuations, and that these phase fluctuations decay at a small enough rate such that they are not simply averaged out from the viewpoint of the mechanical mode. For the largest drive strength, $\alpha_-^{(0)} = 3.6$, the mechanical Wigner distribution $W_m(\beta)$ starts to show signs of bimodality with peaks for a nonzero phase relative to the imaginary axis. The normal mode Wigner distribution $W_c(\alpha_-)$ also starts to look non-Gaussian. However, we attribute the faint ring-shaped structures to finite-size effects due to the Hilbert space truncation.

The Wigner functions shown in Figure \ref{fig:Numerics} are all everywhere positive, which means that they can strictly speaking be interpreted as classical probability distributions. However, it is worth emphasizing that the non-Gaussian fluctuations visible for large probe drives originate from the quantum vacuum noise of the electromagnetic field. These non-Gaussian states could perhaps be useful, e.g., for sensing purposes or as a starting point for generating more interesting states through projective measurements.

While it would be interesting to explore the full nonlinear regime in more detail, a complete analysis is left for future work. We also note that the off-resonant radiation pressure terms in Equation \eqref{eq:HnonlinNormMod} may become relevant as the average phonon number increases beyond $|\beta_\mathrm{crit}|^2$.

\section{Implementation with dispersive optomechanics}
\label{sec:Impl}

The discussion has so far been limited to systems of the type shown in Figure \ref{fig:CSModel} and \ref{fig:Probe}, featuring both coherent scattering from a levitated object and standard dispersive optomechanics. A natural question is whether the effective model we have studied can be realized with large cooperativity ${\cal C}_- \gtrsim 1$ also for purely dispersive optomechanical systems. This would make it relevant to a large variety of experimental platforms, some of which have already reached the regime ${\cal C}_0 > 1$ \cite{Wilson2015Nature,Leijssen2017NatComm,Guo2019,Fogliano2019,Zoepfl2019,Reinhardt2016PRX}. In this Section, we will present a multimode model which shows that it is indeed possible to realize this. However, we do not claim to have found the simplest possible model, such that there may be room for improvement.

\subsection{Two cavity modes}

We have already discussed in Section \ref{sec:EffCoop} how simply having two mechanical oscillators interact with a driven single cavity mode does not work, due to additional mechanical dissipation from off-resonant linear optomechanical interactions. The next natural step is to consider a system with two cavity modes and two mechanical oscillators. It is well-known that such double-cavity systems can realize intermode optomechanical coupling of the form
 \begin{equation}
\label{eq:HTwoCavIdeal}
H_\mathrm{int,2} = \hbar g_{0;2} x_2 \left(a_s^\dagger a_a + a_s^\dagger a_a \right) 
\end{equation}
where photons scatter from one effective mode of the double-cavity system to another when they interact with the mechanical oscillator $b_2$ \cite{Ludwig2012PRL,Stannigel2012PRL,Safavi-Naeini2011NJP,Borkje2012NJP,Paraiso2015PRX,Kharel2019SciAdv}. Ideally, if the other oscillator $b_1$ only interacts with one of the cavity supermodes, say $a_s$,
\begin{equation}
\label{eq:HTwoCavIdeal2}
H_{\mathrm{int,1}s} = \hbar  g_{0;1}  x_1 a_s^\dagger a_s  
\end{equation}
our original interaction Hamiltonians \eqref{eq:Hint1} and \eqref{eq:Hint2} is reproduced by strongly driving the mode $a_a$, such that $a_a \rightarrow \alpha_a$ and $G = g_{0;2} \alpha_a$. However, this would require some very particular system in order to avoid any coupling between modes $b_1$ and $a_a$. While it may be possible, we do not have a particular proposal for how to achieve this. 

If there is indeed optomechanical coupling between modes $b_1$ and $a_a$, , i.e., an additional term $H_{\mathrm{int,1}a} = \hbar  g_{0;1}  x_1 a_a^\dagger a_a  $, it is in fact still possible to realize our effective model with cooperativity ${\cal C}_- \gtrsim 1$ by swapping the roles of $b_1$ and $b_2$, driving both cavity modes, and adjusting the difference between the cavity mode resonance frequencies to match $\omega_{m,1}$. However, for weak coupling $g_{0;2} \ll \kappa$, this would require an extremely large sideband parameter $\omega_{m,2}/\kappa$. Rather than go into detail on this idea, we now proceed to a system with three cavity modes, where this requirement on the sideband parameter can be avoided.

\subsection{Three cavity modes}

\begin{center}
\begin{figure}[htb]
\includegraphics[width=.25\textwidth]{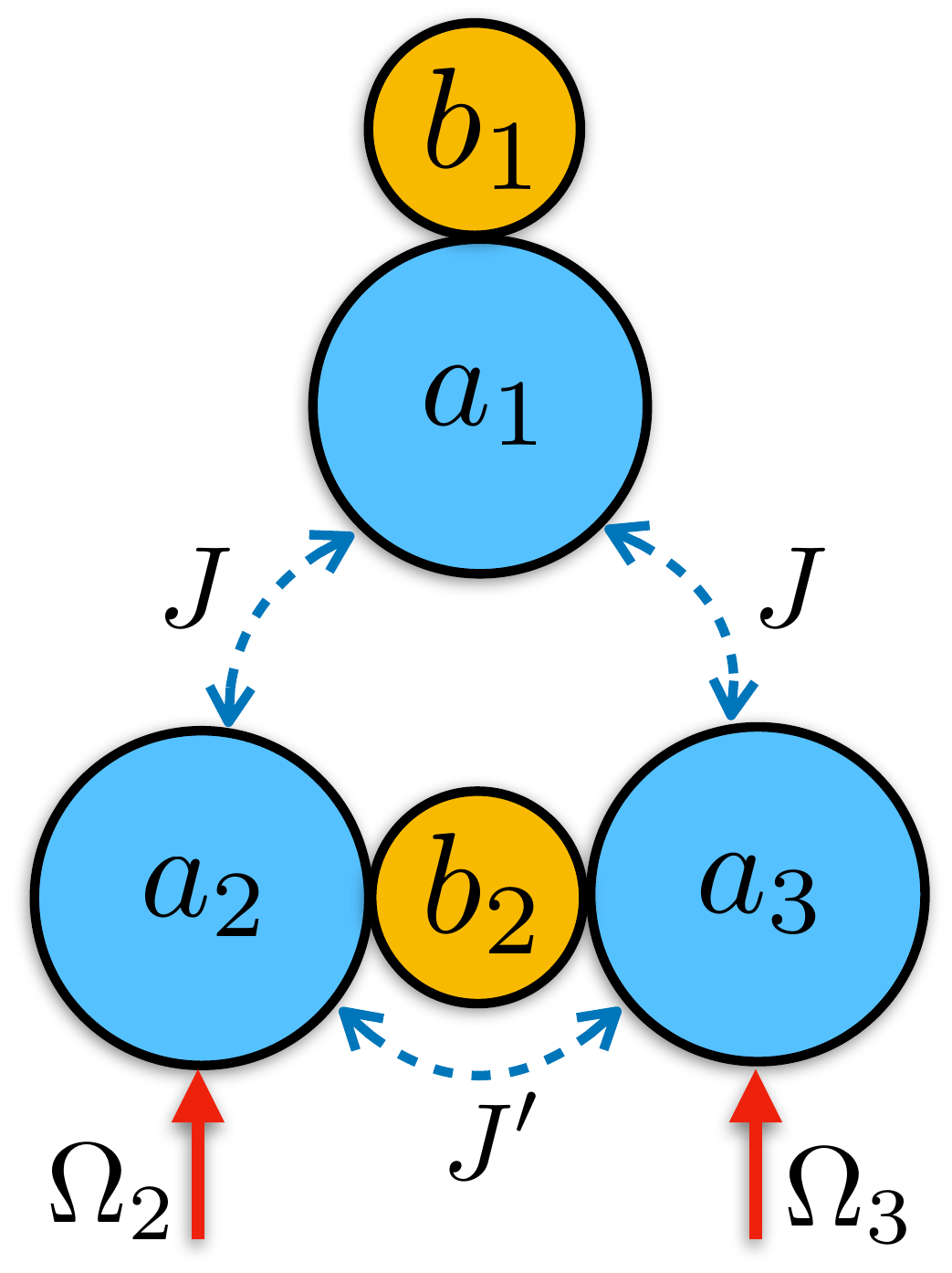}
\caption{Schematic of an optomechanical system with three cavity modes $a_1,a_2,a_3$ and two mechanical modes $b_1,b_2$. The rates $J,J'$ are photon tunneling rates due to direct, bilinear coupling between the cavity modes. The red arrows indicate driving of cavity modes $a_2$ and $a_3$. This system can realize the effective model we have studied if we assume a symmetric setup as described in the text.  }
\label{fig:Setup3}
\end{figure}
\end{center}

We now consider the setup shown schematically in Figure \ref{fig:Setup3}. We let the three cavity modes $a_1, a_2, a_3$ couple to the mechanical modes $b_1,b_2$ according to the interaction Hamiltonian
 \begin{equation}
\label{eq:HThreeCav}
H_\mathrm{int} = \hbar  g_{0;1}  x_1 a_1^\dagger a_1 + \hbar g_{0;2} x_2 \left(a_2^\dagger a_2 - a_3^\dagger a_3 \right)  .
\end{equation}
Note that modes $a_2$ and $a_3$ couple to the same mechanical mode $b_2$, but with opposite signs, a scenario that has been realized both in microwave systems \cite{Lecocq2015PRX} and in the membrane-in-the-middle setup \cite{Sankey2010NatPhys}. Let us also assume that modes $a_2$ and $a_3$ are degenerate, and that they couple bilinearly via photon tunneling at a rate $J'$ larger than the cavity linewidth. This leads to hybridized cavity supermodes  \cite{Ludwig2012PRL,Stannigel2012PRL,Safavi-Naeini2011NJP,Paraiso2015PRX} $a_{s/a} = (a_2 \pm a_3)/\sqrt{2}$, such that the interaction Hamiltonian becomes
 \begin{equation}
\label{eq:HThreeCav2}
H_\mathrm{int} = \hbar  g_{0;1}  x_1 a_1^\dagger a_1 + \hbar g_{0;2} x_2 \left(a_s^\dagger a_a + a_a^\dagger a_s \right)  .
\end{equation}
when written in terms of $a_s$ and $a_a$.

 \begin{center}
\begin{figure}[htb]
\includegraphics[width=.49\textwidth]{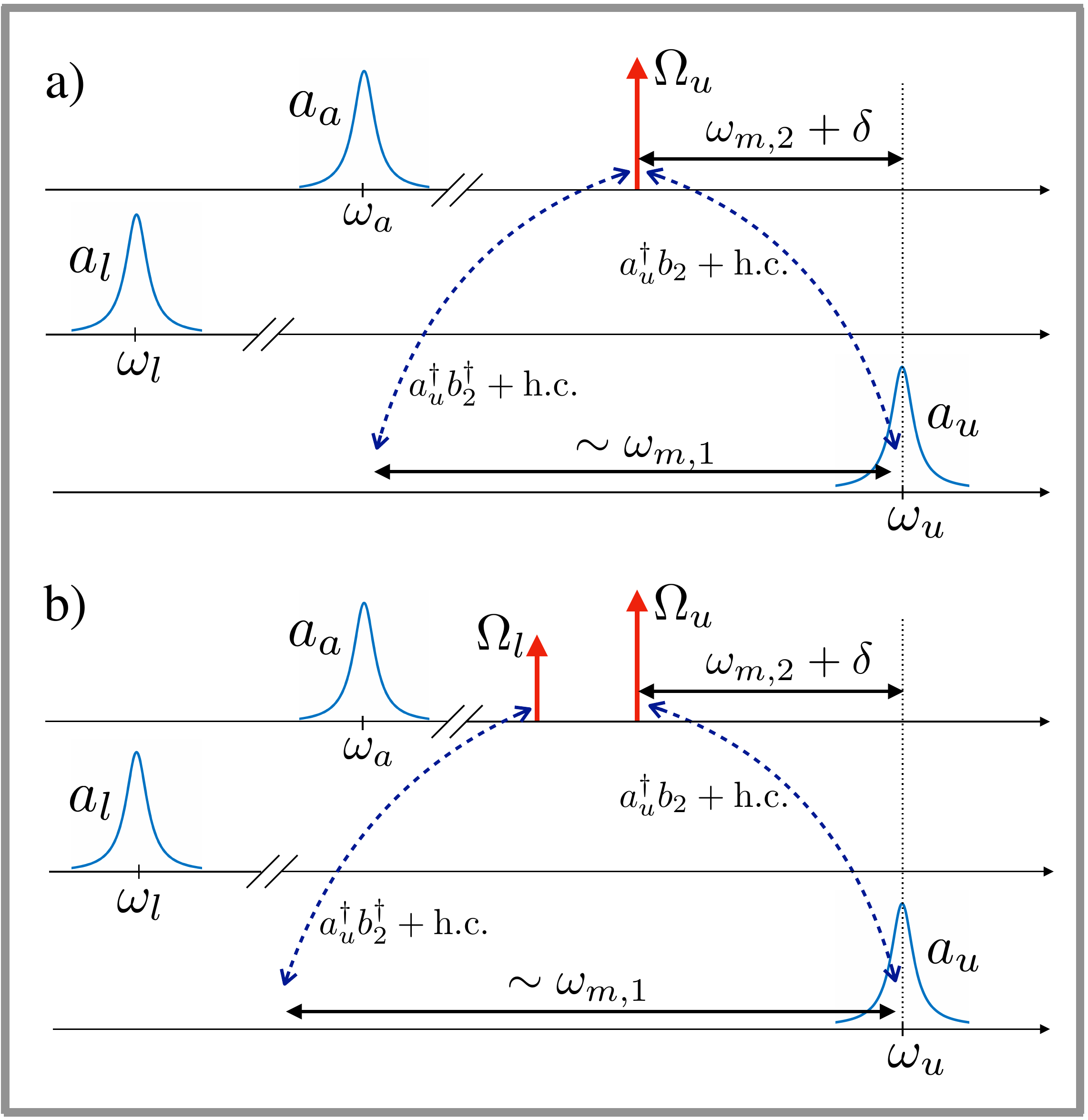}
\caption{The system in Figure \ref{fig:Setup3} is best described in terms of three cavity supermodes $a_u,a_l,a_a$ when the photon tunneling rates $J,J'$ far exceed the cavity decay rates $\kappa$. a) Driving mode $a_a$ at a frequency red-detuned $\omega_{m,2} + \delta$ from the resonance frequency $\omega_u$ recreates our original Hamiltonian from Section \ref{sec:Model1}. b) In this case, it is possible to avoid the requirement \eqref{eq:RoughFreqRel} between the two mechanical oscillator frequencies by using two drive tones.}
\label{fig:Freqs3}
\end{figure}
\end{center}

We assume that cavity mode $a_1$ also couples bilinearly to both modes $a_2$ and $a_3$ in a symmetric setup with equal coupling rates $J$. For $J > \kappa$, cavity mode $a_1$ will then hybridize with the symmetric supermode $a_s$. For simplicity, let us also assume that modes $a_1$ and $a_s$ are degenerate, although this is not actually necessary. This gives a new pair of supermodes $a_{u/l} = (a_1 \pm a_s)/\sqrt{2}$, such that the interaction Hamiltonian can be expressed as
 \begin{align}
\label{eq:HThreeCav3}
H_\mathrm{int} & = \hbar  \frac{g_{0;1}}{2}  x_1 \left(a_u + a_l\right)^\dagger\left(a_u + a_l\right)  \\
& + \hbar \frac{g_{0;2}}{\sqrt{2}} x_2 \left[\left(a_u - a_l\right)^\dagger a_a + a_a^\dagger \left(a_u - a_l\right) \right]  . \notag
\end{align}
We now see that when driving the cavity mode $a_a$ to a large coherent amplitude, which in practice means driving the physical cavities $a_2$ and $a_3$ with opposite phases, we get bilinear optomechanical coupling between mechanical oscillator $b_2$ and modes $a_u$ and $a_l$. However, mode $b_1$ does not couple to mode $a_a$ and thereby does not experience any bilinear interaction.

We now imagine driving mode $a_a$ at a frequency $\omega_{m,2} + \delta$ below the resonance frequency $\omega_u$ of mode $a_u$, as shown in Figure \ref{fig:Freqs3}a). By approximating $a_a$ by its coherent amplitude, we then arrive at exactly the Hamiltonian we started with in Section \ref{sec:Model1} when ignoring mode $a_l$ and renaming $a_u \rightarrow a$. 

In this triple-cavity setup, it is also possible to relax the requirement \eqref{eq:RoughFreqRel} on the relation between the mechanical frequencies. This requires an additional drive tone in mode $a_a$ at a frequency approximately $\omega_{m,1} - (2\omega_{m,2} + \delta)$ below the other drive frequency. Such a configuration is shown in Figure \ref{fig:Freqs3}b). Finally, we note that if one can control the photon tunneling rates $J,J'$, one can reduce the power needed for the drive tones by adjusting the difference between the cavity resonance frequencies such that $\omega_u - \omega_l \approx \omega_{m,1}$ and $\omega_u - \omega_a \approx \omega_{m,2}$. This allows taking advantage of near-resonant virtual processes involving mode $a_l$.

\section{Conclusion and future directions}
\label{sec:Disc}

In this article, we have theoretically studied multimode optomechanical systems and found that, in some cases, the single-photon cooperativity ${\cal C}_0$ can be a genuine figure of merit for observing effects due to intrinsically nonlinear interactions. To our knowledge, this is the first study where this turns out to be the case when simultaneously assuming weak coupling $g_0 \ll \kappa$. The systems we have considered are admittedly very complicated and not immediately accessible in the laboratory. Nevertheless, we hope that our results can inspire new ideas which can help bring cavity optomechanics into the nonlinear regime. 

A straightforward extension of our study would be to consider the nondegenerate parametric oscillator Hamiltonian \eqref{eq:HnonlinNormModNDPO} in the case where both normal modes $c_+$ and $c_-$ are probed with separate drive tones. One would then find a suppression of normal mode coherences similar to what we found in the degenerate case, due to destructive interference. In some sense, the two probe drives would repel each other. In the regime of large effective cooperativity ${\cal C}_{+-} = 4\tilde{g}_{+-}^2/((\kappa_+ + \kappa_-) \gamma_1) \gtrsim 1$, one would again expect photon antibunching in transmission, in the sense that photons would tend to come out at either frequency $\omega_+$ or $\omega_-$, but to a lesser extent at both frequencies simultaneously.

One may also consider coherently driving the system mechanically instead of optically. This leads to a squeezed near-vacuum state in the normal mode in the degenerate case, and a two-mode squeezed state in the nondegenerate case. In the classical theory, a critical point (the squeezing threshold) is again reached for a certain drive strength \cite{Drummond1980OpticaActa}. Above the critical drive strength, the mechanical amplitude stays constant despite increasing the drive and pairs of incoherent photons are emitted from the normal mode(s), reminiscent of the dynamical Casimir effect. For large effective cooperativities, quantum fluctuations will smear out the critical point \cite{Veits1995PRA,Veits1997PRA,Kinsler1995PRA,Drummond2002PRA}. For the degenerate case \eqref{eq:HnonlinNormModDPO}, the squeezed state will gradually pinch into a bimodal distribution, which is interpreted classically as bistability \cite{Drummond1980OpticaActa,Navarrete-Benlloch2014OptExp,Schonburg2015PRA,Jansen2019PRB}.

Other interesting directions could be to investigate the transient behavior of these models, or to study the full nonlinear regime in more detail.

We have seen that the normal modes will always have some nonzero thermal occupation due to the quantum heating effect, even in the absence of thermal phonons. This places a limit on the quantum features one can observe, as discussed in Section \ref{sec:Antibunching} on antibunching. Ideally, we would like to have $n_{\therm,-}$ vanish while still maintaining a sizeable cooperativity ${\cal C}_-$, but this is not possible according to \eqref{eq:CoopLimit}. However, if one were able to engineer the photonic bath in such a way that the density of states vanishes at negative energies $-\omega_\pm$, one could suppress dissipative pair-creation processes and thus the thermal occupation of the normal modes. The removal of processes such as the one shown in Figure \ref{fig:QuantHeat} could also facilitate emission of energy-time entangled photon pairs heralded by single-phonon creation processes if ${\cal C}_- > n_{\therm,1}$. Needless to say, this type of bath engineering is not easy to achieve. An alternative would be to detect all photons both at positive and negative frequencies and postselect on events where there are no photons at negative frequencies. Of course, this would require very large detection efficiencies and the ability to detect negative and positive frequencies separately through filtering.

\acknowledgments{
The author acknowledges useful discussions with Stefan Walter and Francesco Massel. The numerical calculations were performed using the software QuTiP \cite{Johansson2012CPC,Johansson2013CPC}. This work was performed with financial support from the Research Council of Norway (Grant No. 285616) through participation in the QuantERA ERA-NET Cofund in Quantum Technologies (project QuaSeRT) implemented within the European Union's Horizon 2020 Programme. 
}


%

\end{document}